\documentclass[useAMS,usenatbib]{mn2e}

\usepackage{graphicx}
\usepackage{amssymb}
\usepackage{natbib}
\usepackage{threeparttable}
\newcommand{\eqb}{\begin{eqnarray}}
\newcommand{\eqe}{\end{eqnarray}}
\newcommand{\sth}{\sigma_{\rmn T}}
\newcommand{\gpmn}{\gamma_{\rm p,min}}
\newcommand{\gpmx}{\gamma_{\rm p,max}}
\newcommand{\gemn}{\gamma_{\rm e,min}}
\newcommand{\gemx}{\gamma_{\rm e,max}}
\newcommand{\tpesc}{t_{\rm p,esc}}
\newcommand{\teesc}{t_{\rm e,esc}}
\newcommand{\tecool}{t_{\rm e, cool}}
\newcommand{\mpr}{m_{\rm p}}
\newcommand{\me}{m_{\rm e}}

\newcommand{\ub}{u_{\rm B}}
\newcommand{\lpinj}{\ell_{\rm p}^{\rm inj}}
\newcommand{\leinj}{\ell_{\rm e}^{\rm inj}}

\newcommand{\tcr}{t_{\rm cr}}

\title[Variability signatures of leptohadronic systems]
{Mrk 421 as a case study for TeV and X-ray variability
in leptohadronic models}
\author[A. Mastichiadis, M. Petropoulou and S. Dimitrakoudis]{A. Mastichiadis (AM) \thanks{E-mail:
amastich@phys.uoa.gr}
M. Petropoulou (MP) \thanks{E-mail:
maroulaaki@gmail.com } and
S. Dimitrakoudis (SD) 
\thanks{E-mail: sdimis@phys.uoa.gr}
 \\
Department of Physics, University of Athens, Panepistimiopolis, GR 15783 Zografos, Greece}
\begin{document}
\date{Received.../Accepted...}

\pagerange{\pageref{firstpage}--\pageref{lastpage}} \pubyear{2013}

\maketitle

\label{firstpage}
\begin{abstract}
We investigate the origin of high-energy emission in blazars
within the context of the leptohadronic one-zone model.
We find that $\gamma$-ray emission can be attributed to synchrotron radiation either
from protons or from secondary leptons produced
via photohadronic processes.
These possibilities imply 
differences not only in the spectral energy distribution (SED)
but also in the 
variability signatures, especially in  the X- and $\gamma$-ray regime.
Thus, the temporal behavior of each leptohadronic scenario can be used
to probe the particle population responsible for the high-energy emission
as it can give extra information not available by spectral fits.
In the present work
we apply these ideas to the non-thermal emission of Mrk 421, 
which is one of the best monitored TeV blazars. 
We focus on the observations of March 2001, since during  that period 
Mrk 421 showed multiple flares that have been observed
in detail both in X-rays and $\gamma$-rays. First, we obtain
 pre-flaring  fits to the SED using the different types of leptohadronic
 scenarios.
Then, we introduce random-walk type, small-amplitude variations on the
injection compactness
or on the maximum energy of radiating particles and follow
the subsequent 
response of the radiated photon spectrum.
For each leptohadronic scenario, we calculate
the X-ray and $\gamma$-ray fluxes 
and investigate their possible correlation. 
Whenever the `input'
variations lead, apart from flux variability, also to spectral variability,
we present the resulting relations between the  spectral index and the flux, both 
in X-rays and $\gamma$-rays. 
We find that proton synchrotron models are favoured energetically
but require fine tuning between electron and proton parameters
to reproduce the observed quadratic behaviour between X-rays and TeV
$\gamma-$rays. On the other hand,
models based on pion-decay can reproduce this
behaviour in a much more natural way.

\end{abstract}

\begin{keywords}
astroparticle physics -- radiation mechanisms: non-thermal -- gamma rays: galaxies -- galaxies: active
-- BL Lacertae objects: general
\end{keywords}

\section{Introduction}

Blazars are a subclass of Active Galactic Nuclei (AGN)
with a non-thermal emission covering most
of the electromagnetic spectrum, i.e. from radio up
to high-energy gamma-rays. Their broadband emission, that
originates from a relativistic jet oriented close to 
the line of sight, is Doppler boosted and shows
no evidence of spectral lines.
The spectral energy distribution
(SED) of TeV-emitting blazars consists of two smooth,
broad components (e.g. \citealt{ulrichetal97, fossatietal98}). The first one extends from
the radio up to the X-rays with a peak in
the soft or hard X-rays, while the second one
extends up to TeV energies, with a peak energy around
$0.1$ TeV, although this is not always clear \citep{abdoetal11}. 

Although the lower energy bump is attributed to
the synchrotron radiation
of relativistic electrons, the origin of the
high-energy component is still under debate.
Theoretical models
are divided into two classes: (i) leptonic and
(ii) (lepto)hadronic, according to the type of particle responsible
for the $\gamma$-ray emission.
In the leptonic scenario, the high-energy component 
is the result of Compton scattering of electrons
on a photon field; 
 the seed photons  can  come either from
synchrotron emission of the same electron population
(SSC models) (e.g. \citealt{maraschietal92, konopelkoetal03}) or 
from an `external' region, i.e. from the accretion disk 
\citep{dermeretal92,dermerschlickeiser93} or from the broad line
region \citep{sikoraetal94, ghisellinimadau96, boettcherdermer98} (EC models). 
In the hadronic scenario, on the other hand, 
the $\gamma$-ray emission is the result of
proton or secondary lepton (through photohadronic interactions)
 synchrotron radiation \citep{mannheimbiermann92, aharonian00, mueckeetal03}.
For a recent review see \cite{boettcher12}.

The above models have been successfully applied
for fitting the overall SED of TeV blazars by assuming stationary
conditions, at least in most cases -- this is particularly true for
the class of hadronic models. 
However, one major feature of blazar emission is the 
variability observed in almost all energies (see for example \citealt{raiterietal12}), 
which further implies that stationary conditions might not
apply. Moreover, variability can 
provide additional constraints on source modelling and therefore it 
can be used to lift the apparent degeneracy between leptonic
and hadronic models. 
As the above models use very different processes
of widely varying cooling timescales,
%Monochromatic emission in the context of the above models
%is produced by particles of different energy and/or type, 
%different cooling
%timescales, 
%through interactions with different cross sections and 
%`targets'. 
one expects that the system will react
in a different way to variations on one or more source
parameters, such as the injection rate of fresh particles.
For example, one of the successes of SSC leptonic modelling 
is that it can reproduce the quadratic behaviour between the X-ray
and TeV fluxes \citep{mastkirk97, krawczynskietal02}.
The lack, thus far, of time-dependent hadronic models, has forbidden
an analogous study of their expected radiative signatures.

Recently \cite{DMPR2012} have presented an one-zone 
hadronic model which is time-dependent.
In the present paper we make use of the numerical 
code presented there in search for variability
signatures in the context of these models. 
As an illustrative example,
we have chosen to focus on
the relatively recent, high quality
observations of Mrk421 \cite{Fossati2008}, that provide us with
 both spectral and temporal information. 
While we do not attempt to make a detailed spectral 
or temporal fit to the observations,
we use these as a springboard to examine the 
trends expected within the hadronic models.

The present paper is organized as follows: In \S2
we present the basic principles of the
one-zone hadronic model and we
comment on the
possible options for fitting
the SED of a blazar; the type of
variations adopted in our simulations 
is also presented.
In \S3 we present our results starting with
the multiwavelength (MW) fits obtained using three different (lepto)hadronic
models. In subsections 3.2 - 3.3 we present
the variability signatures obtained for each model caused by
variations that we have applied on the injection
compactness or on the maximum energy of particles respectively. 
We conclude in \S4 with a summary and a discussion of our
results.
For the required transformations
between the reference systems of the blazar and the observer, 
 we have adopted a cosmology 
with $\Omega_{\rm m}=0.3$, $\Omega_{\Lambda}=0.7$ and 
$H_0=70$ km s$^{-1}$ Mpc$^{-1}$, where the redshift of Mrk 421 $z=0.031$ 
corresponds to a luminosity distance $D_{\rm L}=0.135$ Gpc.

\section{The model}

\subsection{General Principles}

In what follows we use the one-zone hadronic model as described in 
%\citet[henceforth DMPR]{DMPR2012}.
\cite{DMPR2012} -- henceforth DMPR.
For completeness reasons we repeat here its basic points.
We  consider
 a spherical blob of radius $R$ moving with a Doppler factor
$\delta$ with respect to us and containing a magnetic field of strength $B$. 
We further assume that ultra-relativistic protons with a power law distribution of index $p_{\rm p}$
between some energy limits  $\gpmn$ and $\gpmx$ are 
injected into the source (we will be using Lorentz factors to denote proton or electron energies throughout this paper). This injection can be characterised
by a compactness
\eqb
\lpinj={{L_{\rm p} \sth}\over{4\pi R \mpr c^3}}
\label{lpinj}
\eqe
where $L_{\rm p}$ is the proton injected luminosity
and $\sth$ is the Thomson cross section.

Protons can lose energy via three channels:
(a) synchrotron radiation, (b) photopair (Bethe-Heitler) and 
(c) photopion production. 
The effect that any of those three processes has on the proton distribution function 
depends on the specific parameters of the system, therefore all three have 
to be taken into account in a kinetic equation that also includes a proton injection and a proton escape term. Furthermore, since
the above processes will create photons and other secondary particles
 (which will eventually decay to electrons and positrons, both of which we will hereafter refer to as electrons), one has to also follow the evolution of photons and electrons, by writing two additional kinetic equations for them
\footnote{Neutrons and neutrinos are also byproducts of photopionic
interactions and one therefore should, in principle, write two more
equations for them -- see DMPR. For our present case , however,
one can safely ignore them.}.
Assuming that the particles have a uniform distribution inside the source,
one can write a system of partial, with respect to time and energy,
integrodifferential equations, 
whose solution  gives the corresponding particle distribution in addition to the 
multiwavelenrth photon spectrum emerging from the source. 

The total number of free parameters used in this case is eight: 
The radius $R$ of the source,
the magnetic field strength $B$,  the proton
injection compactness $\lpinj$, the lower and upper 
 Lorentz factors of the proton power law distribution $\gpmn$ and $\gpmx$, as well as the 
proton index $p_{\rm p}$ and the escape time from the source $\tpesc$.
To these one should add  
the Doppler factor $\delta$ of the emitting blob, which is used
to boost the radiation and convert the source-frame quantities
into observed flux. 

In the case where electrons are injected, in addition to protons,
one could introduce
the corresponding electron parameters which are 
their injection compactness $\leinj$,
which is related to their luminosity $L_{\rm e}$ in the same way
 as described in eq. (\ref{lpinj}) with $\me$ replacing $\mpr$, 
% 
% through the relation
% \eqb
% \leinj={{L_e\sigma_T}\over{4\pi Rm_ec^3}},
% \eqe
 the upper and lower cutoff of their injected spectrum $\gemn$ and $\gemx$ respectively,
their slope
$p_{\rm e}$ and their escape timescale $\teesc$.
In order to  reduce the number of free parameters one can assume that
$\gpmn=\gemn=1$ and $\tpesc=\teesc$.
%$p_{\rm p}=p_{\rm e}=p$ 

Depending on the assumptions made about the time dependence
of the parameters, the above scheme 
can be used to derive both steady-state and time-dependent solutions.
Thus, if all parameters are constant in time the system will 
eventually reach a 
steady-state --  note however that hadronic plasmas can become 
supercritical and in such cases they can exhibit limit cycle behaviour
\citep{PM12b} -- hereafter PM12b.
If, on the other hand, the system is in the subcritical regime and 
we allow for one or more parameters to have some explicit time
dependence, then the system will not reach a steady state but it
will show continuous temporal
variations, which will reflect the corresponding ones imposed on the
input parameters and will have an impact on the produced spectrum.

Therefore, one can first use the numerical code to obtain the SED of a source
in a stationary state and then introduce perturbations in one or more of the
fitting parameters to check the variability patterns in the MW spectrum. 
Similar methods have been applied in the case of leptonic models by
\cite{mastkirk97} and \cite{krawczynskietal02}. However, in the framework of hadronic modelling 
we find that
there are different combinations of radiative processes which can,
in principle, give acceptable fits to the SED of blazars. Thus, 
it is possible that each fitting combination will have distinct
temporal signatures which are worth investigating. 
In what follows
we make a qualitative discussion on the various options one has 
of fitting the SED of blazars using a hadronic model and then present our
method regarding temporal variations.

\subsection{Fitting the SED}

In contrast to the leptonic model where the relevant radiative processes 
are few, the hadronic model involves many processes which can make the
radiated spectrum quite complicated. However,
as it was shown in DMPR, there are certain limiting, yet intuitive, cases,
where the derived spectrum has a particularly simple form.  
Thus, in the  case where all features in the MW spectrum can be attributed to protons\footnote{A primary leptonic component may also exist, 
provided its contribution to the synchrotron emission
is much smaller than that of protons -- see Appendix A for more details.},
% or 
% $\leinj  << \frac{4 \sth R u_{\rm B}}{3 \me c^2}\lpinj \gpmx \left(\frac{\me}{\mpr} \right)^3 \frac{2-p}{3-p}$, for $p<2$.
%For the parameters of model `H' in Table \ref{table1} the condition becomes $\leinj << 10^{-8}$.}  --  
%assuming $\gamma_{\rm e,max} = 3 \times 10^4$, 
a case we will refer to as `pure hadronic' or simply `model H', the spectrum
has four distinctive features. One due to proton synchrotron radiation,
two due to synchrotron radiation of electrons produced respectively in photopair and
charged pion decay and one due to $\gamma-$rays from
neutral pion decay. What is also
interesting is that the frequencies where the 
three first peaks occur have a fixed ratio
between them and scale as $(m_e/m_p):1:\eta^2_{pe}$ where the value
$\eta_{pe}\simeq 150$ is deduced empirically from the results of the SOPHIA code \citep{SOPHIA2000}.
So the frequencies of the first and third peaks 
are about eight orders of magnitude apart and therefore if the   
first one happens to be in the
X-ray regime then, interestingly enough, the third one will be at TeV 
$\gamma-$rays\footnote{Note that, as pointed by PM12b, the electromagnetic cascade that ensues
from the absorption of the $\pi^0-$decay $\gamma-$rays also contribute to the same
energy regime.}. 
Clearly this coincidence needs some investigation as far as MW modelling
is concerned.

The introduction of a high luminosity leptonic component can 
change the picture significantly.  
If electrons have suitable parameters as to 
explain the X-rays, then there are two different options for fitting the
TeV $\gamma$-rays with protons. 
The first is from the process described above, i.e. $\gamma-$rays
are produced from a combination of the
synchrotron radiation of electrons produced in charged pion decay
and of the electromagnetic cascade induced from neutral pion decay.  
We shall refer to this mechanism for TeV $\gamma-$ray production as
`pion induced' and we will denote the model with the acronym `LH$\pi$'.
As it turns out this model  requires
proton Lorentz factors not higher than $10^7$ 
and intermediate magnetic field strengths ($\sim 10$ G). 
The same holds, in general, for model H.

The second option is to fit the SED 
with very high proton energies ($\gpmx\sim 10^{10}$) and high magnetic
fields ($> 20$ G). In the latter case, proton synchrotron emission is
responsible for the TeV emission -- this is the most popular 
fitting method of the relevant observations and we will refer to it
as the `proton-synchrotron' case or as `LHs model'. Finally, both
LH$\pi$ and LHs models fall into the more general category of  `leptohadronic' cases.

The above picture holds for low enough compactnesses. For higher ones
photon-photon absorption, especially of the $\pi^0$-component that always
peaks at very high frequencies, begins to dominate.
However, as it was discussed in PM12b, the absorbed energy will
be reemitted mostly at frequencies close to the peak
produced by the radiation of charged pion-produced electrons, 
and only if the compactness becomes relatively high will the photon-photon absorption process become severe enough to distort the spectrum through electromagmetic
cascades and redistribution to lower frequencies. At still higher compactnesses the system undergoes 
a phase-transition and it becomes supercritical: various radiative loops
such as photopair-synchrotron \citep{kirkmasti92}
 and photon quenching \citep{stawarzkirk, petromasti11} rapidly take
energy away from the protons and redistribute it to electrons and radiation.
In this case the system becomes very efficient, since a large fraction
of the input proton luminosity is turned into radiation; however
this is done at the cost of the spectral features described earlier which are
destroyed by a very strong, non-linear photon-photon absorption. 
Clearly one should seek successful spectral fits of MW blazar
observations while the system is in the subcritical regime;
 note however that one can use supercriticalities 
in various contexts \citep{mastikaza06,mastikaza09,PM12a}.   

%\subsection{Variability}

\subsection{An algorithm for inducing time variability}

DMPR have examined some examples of variability in 
the case of a pure hadronic model and showed that 
the system behaves like a SSC leptonic one, 
in the sense that proton synchrotron radiation produces the soft photons
which act as targets for the 
 photopair and photopion processes. Thus, in complete analogy to the leptonic SSC,
hadrons interact with their own radiation. Consequently, one expects that variations 
in the proton injection parameter will produce a quadratic relation 
between the  proton synchrotron and the pion induced components. 
%DMPR have also shown that there are parameter regimes 
%where the two aforementioned components vary cubically but the required 
%parameters fail to reproduce the SED of blazars.    

While the inclusion of primary electrons introduces more free 
parameters in the model and facilitates the spectral fitting, it 
complicates the 
problem of inducing variability since now one should
introduce two more free parameters, one of which relates
the amplitude of variations between electrons and protons while
the other relates their phases.  
In other words, while the
pure hadronic model is expected to have a well-predicted temporal behaviour, 
since variations can be induced by changing a single parameter,
the leptohadronic models will have, by neccessity,  more complicated 
temporal patterns. 

In order to simulate temporal variations we will adopt the following algorithm:
\begin{enumerate}
 \item First we will obtain fits to 
a stationary/low state SED of a TeV blazar
with each of the three aforementioned models (H, LH$\pi$, LHs).
\item We will then introduce variations on some key parameter
(injection luminosity or maximum energy of particles) and follow
in time the changes
that these perturbations introduce to the SED of the source.
\item Since most observations focus on the correlation between
X-rays and TeV $\gamma$-rays, we will focus on these energy bands and 
find the correlations expected in each of these three different
scenarios. 
\end{enumerate}

For concreteness, we will apply this approach to the observations
of Mrk 421 by \cite{Fossati2008}. These sessions have produced a wealth of good quality, 
time-dependent data both at the X-ray (\textit{RXTE}) and TeV (H.E.S.S.) regimes.
However, we should warn the reader that we do not use the above data 
in order to make detailed fits but rather as a basis for 
comparing the general trends of our simulations with respect to them.
Specifically, in the first step of the algorithm described above, 
we do not attempt a very detailed fitting of the SED but
we use the approach of \cite{PM12a}, where a family of possible fits with low reduced
$\chi^2$ (below 1.5) 
can be considered as acceptable.\footnote{Clearly introducing
more free  parameters would have improved the fit but this would have 
acted against the scope of the present paper.} 
At any rate, it turns out
that the expected temporal variations presented in \S 3 are largely independent of the
value of the $\chi^2$ of the particular fit. 
% \begin{figure}
% \centering
%  \resizebox{\hsize}{!}{\includegraphics{./figs/pdsinp.eps}}
% \caption{Spectral power density of the temporal variations used in 
% our simulations (black line). The results of linear fit (grey line) are shown in the inset legend.}
% \label{spd}
% \end{figure}

Motivated by the results of long-term variability studies of Mrk 421 and other 
prototype blazars (e.g. \cite{emmanoulopoulosetal10}),  we have introduced, for the temporal variations, a random-walk type
of change in one of the fitting parameters.
Thus, we use 
\eqb
a_{i+1}=a_i+(-1)^{\rm int(\xi)},\quad i=0,1,2,...
\eqe
where $\xi$ is a uniformly distributed random number in the range (0,10).
%The  power spectral density (PSD) of the timeseries $\{a_i\}$ (black line)
%along with a linear fit (grey line) are shown in Fig.~\ref{spd}. The derived slope
%is $\simeq -1.9$ in agreement with the quadratic slope 
%of the PSD of a brownian noise.
The integer parameter $a_i$  is then scaled in such a way as to produce
the following change in the parameter $y$ we wish to vary
\eqb
y_i=y_0(1+fa_i),\quad i=1,2,...
\label{var}
\eqe
where $y_i$ is the value of the parameter at time $t_i$ and
$f$ a multiplication factor -- in all the examples shown
in the present paper we have chosen $f=0.05$, i.e. the 
parameter in question varies only by 5\% between crossing
times. For the values of $y$ between two successive crossing times, 
we have chosen a linear interpolation scheme.
Obviously $a_0=0$, in order for the initial value
of the parameter in question to match its corresponding value
in the steady state fit $y_0$.

\section{Results}

We proceed next to show some characteristic results. We will begin by showing
the spectral fits obtained and then we will focus on the variability
induced on the models by varying first the injection compactness ($\ell^{\rm inj}$) 
and then the maximum energy ($\gamma^{\rm max}$) of particles.

\subsection{Spectral fits}
\begin{figure}
  \centering
 \begin{tabular}{l}
\includegraphics[width=7.7cm, height=7.cm]{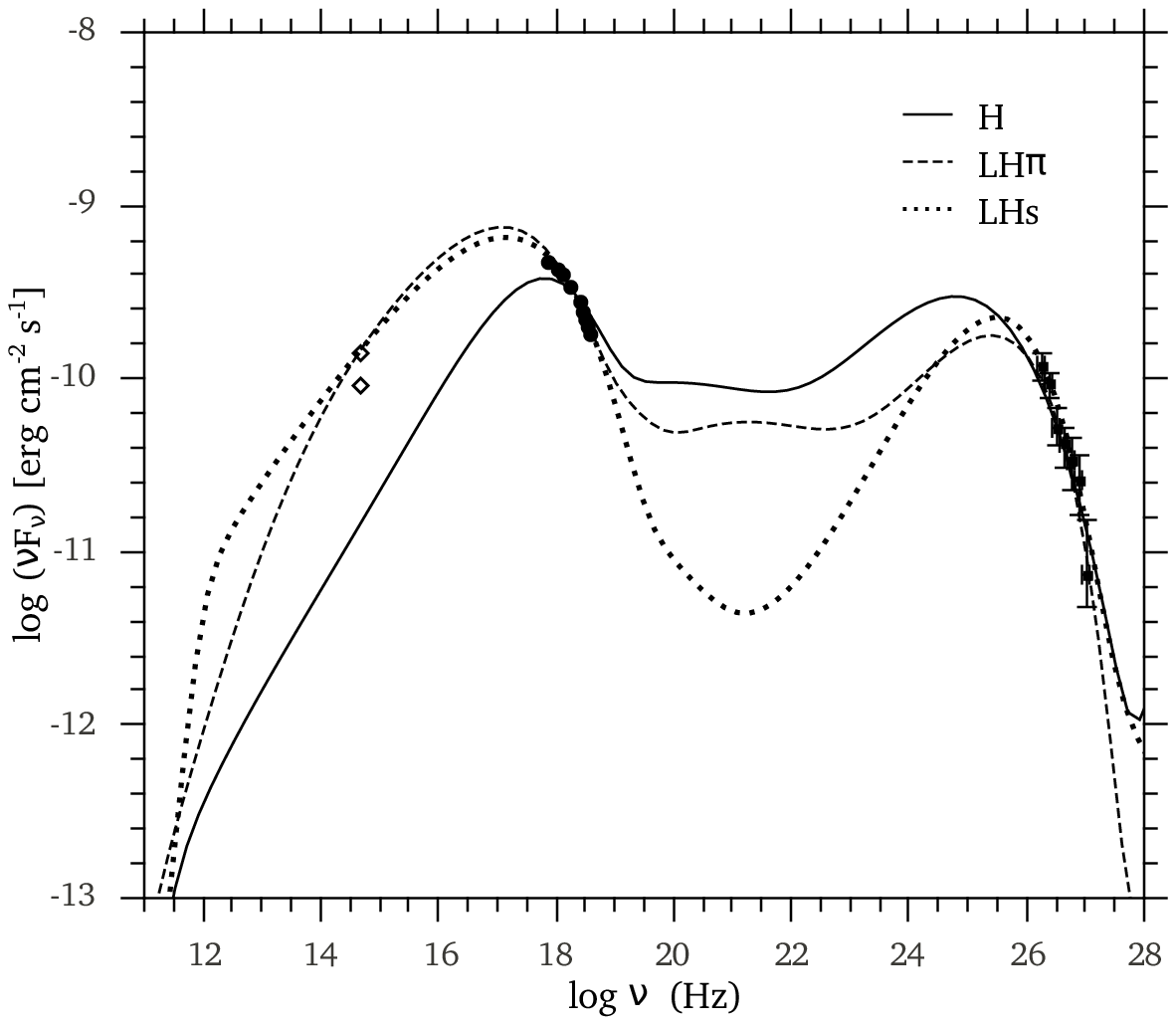} \\
\includegraphics[width=7.cm, height=7.cm]{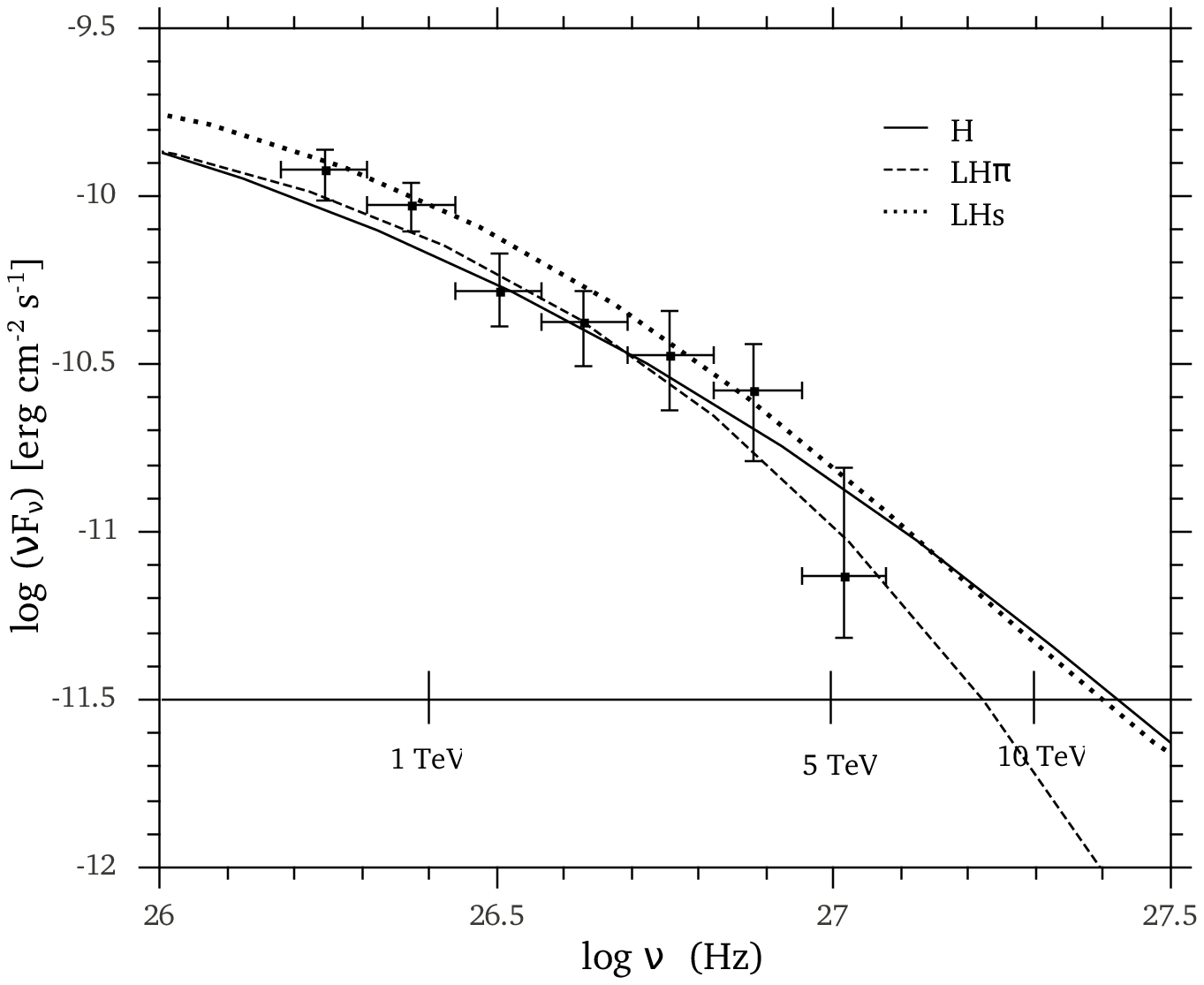} 
  \end{tabular}
\caption{Top panel: Multiwavelength fit to the March 22nd/23rd data of Mrk 421
using (a) a purely hadronic one-zone model (solid line), (b) a leptohadronic one-zone model where the TeV component of the SED is produced by synchrotron 
radiation of electrons resulting from photohadronic interactions (dashed line), and (c) as before but with the TeV component 
of the SED produced by proton synchrotron radiation (dotted line). For the parameter
 used see Table \ref{table1} .
%The parameters used for each model are: (a) 
% $R=2.4\times 10^{15}$ cm, $B=20$ G, $\gpmx=6.3\times 10^5$, 
% $p=1.2$, $\lpinj=0.012$ and $\tpesc=1$, (b) , (c)  {\bf (fill in later)}. 
Bottom panel: The above zoomed in on the TeV energy range.}
\label{MW-all}
\end{figure}
Figure \ref{MW-all} (top panel) shows the spectral fits we obtained using the three models
described in the previous section
 to the pre-flaring state of Mrk 421 on the night of 22nd/23rd March 2001.
All data points used for the fitting are contemporaneous;
for more details on the observations
see \cite{Fossati2008}.  
Different types of lines correspond to different models as shown in the plot.
A zoom in the TeV energy range is also shown in the bottom panel
of the same figure. Note that, in all three models, the TeV $\gamma$-rays are fitted with
the cutoff of the proton (LHs) or secondary lepton (H, LH$\pi$) synchrotron spectrum; 
in the latter cases (H and LH$\pi$), synchrotron radiation from Bethe-Heitler
 electrons produce a distinctive broad spectral feature
which lies between the proton synchrotron and the pion induced components.
The fitting parameters of each model are listed in Table \ref{table1}.
In addition, the comoving energy densities of protons ($u_{\rm p}$), electrons ($u_{\rm e}$) and
photons ($u_{\gamma}$) at the steady state are listed. The observed luminosity
of the jet has been calculated according to
\eqb
P^{\rm obs}_{\rm jet} \approx \pi R^2 \delta^2 \beta c (u_{\rm B}+u_{\rm p}+u_{\rm e}+u_{\gamma}).
\eqe
Although all three models can give us acceptable fits to the SED,
 the pure hadronic is
the most energy demanding, having the largest 
ratio of particle to magnetic energy density ($u_{\rm p}/u_{\rm B} \approx 2\times10^3$) and
the highest jet luminosity. Clearly, as far as energy requirements go, the proton synchrotron
model is, by far, the most economic of the three.
\begin{table}
\centering
\caption{Parameters for the pure hadronic and leptohadronic fits to the pre-flaring
state of Mrk 421 (see Fig.~\ref{MW-all}).}
\label{table1}
\begin{threeparttable}[b]
\begin{tabular}{l  c c c}
 \hline 
Parameter symbol  &  Model H  & Model LH$\pi$  & Model LHs \\
%\phantom{} &  & (`LH$\pi$') & (`LHs') \\
\hline \hline
 $R$ (cm) & $3.2 \times 10^{15}$ & $3.2 \times 10^{15}$ & $3.2 \times 10^{15}$ \\
 B (G) &  20    & 5 & 50 \\
$u_{\rm B}$ (erg cm$^{-3}$) & 15.9 & 1.0 & 99.5 \\
 $\delta$ & 16   &  31 & 21 \\
$t_{\rm var}^{\rm obs}$ (hr) & 1.8 & 0.9 & 1.4  \\ 
\hline
  $\gpmx$ & $8\times 10^5$  & $4\times10^6$ & $4 \times 10^9$ \\
  $p_{\rm p}$ & 1.3 & 1.5 & 1.5 \\
  $\lpinj$ &$1.6\times10^{-2}$ &  $7.9\times 10^{-4}$ & $1.6 \times 10^{-7}$ \\
\hline
 $\gemx$ & -- & $3\times10^4$ & $8 \times 10^3$ \\
 $p_{\rm e}$ & -- & 0.7 & 0.5 \\
$\leinj$ & -- &  $2\times 10^{-5}$ & $5 \times 10^{-5}$ \\
\hline
 $u_{\rm p}$ (erg cm$^{-3}$)\tnote{1} &    $3.2\times10^4$ & $1.6\times10^3 $ & $2.9\times10^{-1}$ \\
$u_{\rm e}$ (erg cm$^{-3})$ & $2.3\times10^{-4}$   & $2\times10^{-3}$ & $3.4\times10^{-3}$     \\
$u_{\gamma}$ (erg cm$^{-3}$) & 1.2 & 0.1 & $3.6\times10^{-1}$ \\
\hline
$P^{\rm obs}_{\rm jet}$ (erg/s) & $6.9\times10^{48}$ & $1.3\times 10^{48}$& $2.4\times10^{44}$   \\
\hline
\end{tabular}

 \begin{tablenotes}
    \item[1] The listed 
    particle and photon energy densities correspond to the pre-flaring fit, i.e. to a steady state of the system.
  \end{tablenotes}
 \end{threeparttable}
\end{table}

\subsection{Varying $\lpinj$}
\begin{figure}
\begin{tabular}{c}
\includegraphics[width=7.2cm, height=7.2cm]{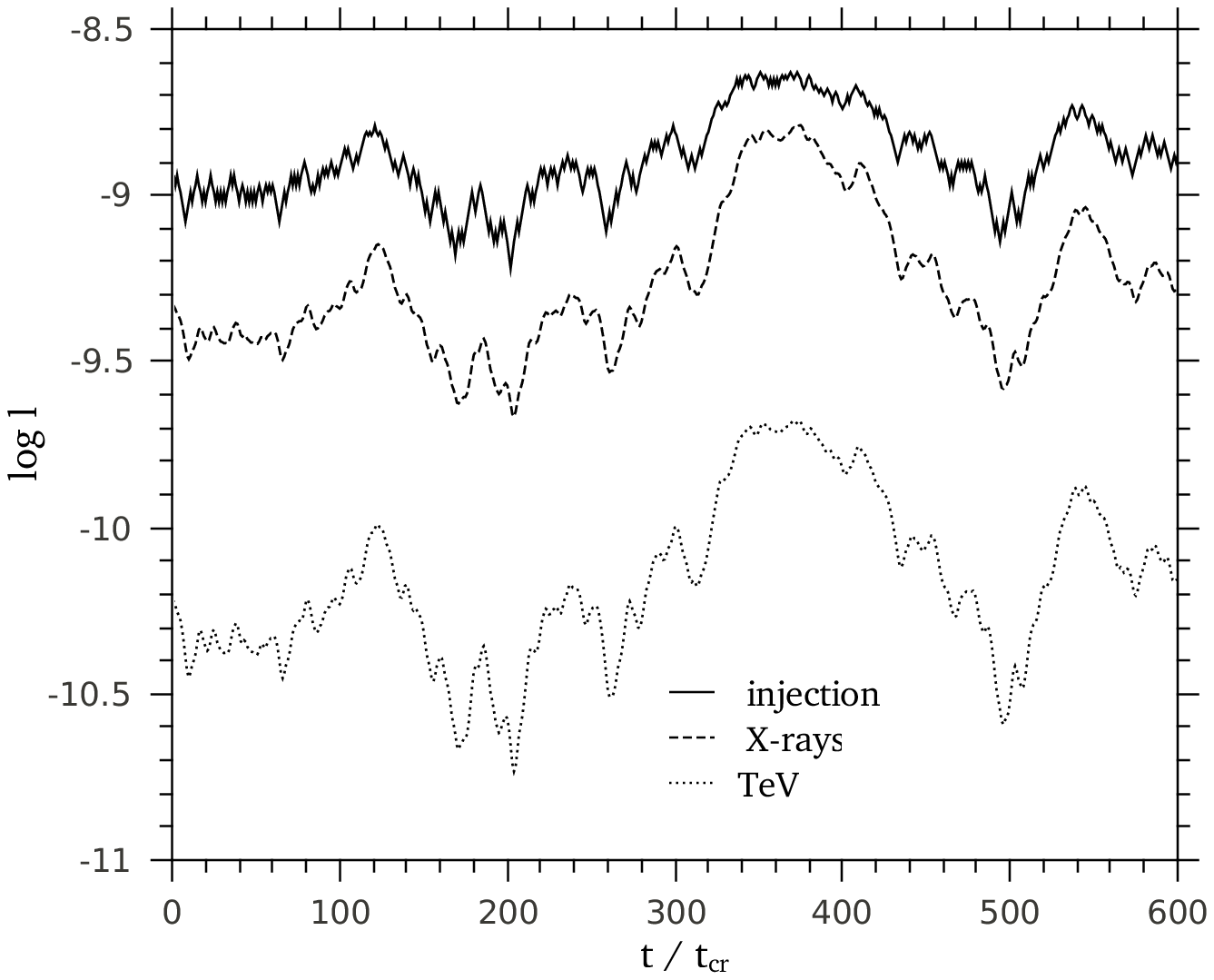}\\
\includegraphics[width=7cm, height=7cm]{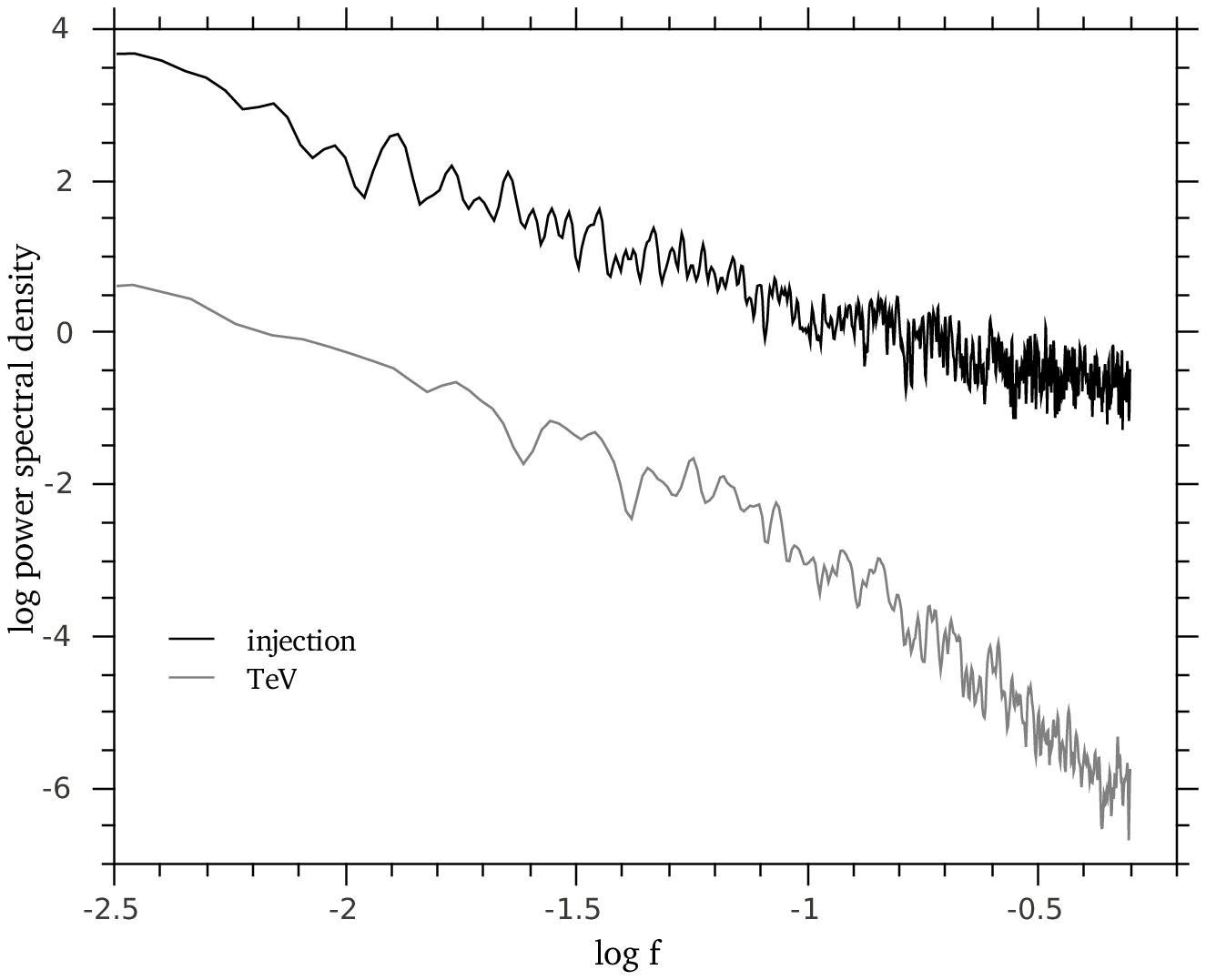}
%\resizebox{\hsize}{!}{\includegraphics{./figs/fig2.eps}}\\
%\resizebox{\hsize}{!}{\includegraphics{./figs/psd_hadro_all.eps}}
\end{tabular}
\caption{Top panel: X-ray (dashed line) and TeV (dotted line) lightcurves 
for the pure hadronic fit shown in Fig.~\ref{MW-all}, obtained by varying $\lpinj$ 
(thick solid line); the latter is shifted downwards by 7 units in the y-axis for clarity reasons.
%For the adopted parameters, $\tcr=1,15$ hrs.
For the adopted parameters, one $\tcr$ corresponds to $\simeq 1.8$ hrs in the observer's frame. 
Bottom panel: Power spectral density (PSD) of the injection time series (black line) and of the derived
TeV light curve (grey line). }
\label{lc-hadronic}
\end{figure}

One straightforward way of producing flaring activity is to vary
the injection luminosity of the high-energy radiating particles. Thus,
we start with the pure hadronic case and choose the proton injected compactness $\lpinj$ to 
be the varying parameter.
%The first parameter we will be varying is the proton injected compactness $\lpinj$. 
Top panel of Fig.~\ref{lc-hadronic} depicts the X-ray (middle)
and the TeV lightcurves (bottom) 
centered at $10^{18}$ and $2.5\times10^{26}$ Hz respectively. We note that, in what follows, any reference
to the TeV and X-ray fluxes will mean the integrated fluxes in the ranges $(1.6\times 10^{26}-10^{27})$ 
Hz and $(7.2\times10^{17}-3.6\times10^{18})$ Hz respectively. 
For comparison reasons, the 
injected proton luminosity (top curve) is also plotted after  being rescaled.

In the bottom panel of the same figure, the power spectral densities (PSD)
of the injection and TeV light curves are shown with black and grey lines respectively. 
The corresponding PSD for the X-ray light curve is not shown, since
it is similar to that of the TeVs apart from a normalization factor.
While the PSD of the injection time-series can be fitted by a power-law with slope $\sim -1.9$,
in agreement with the characteristic $-2$ slope of brownian noise, the PSD
of the TeV light curve 
is best fitted with a broken power-law having the same slope with the injection's
PSD at low frequencies (large timescales) but steepens significantly at higher frequencies
 ($f \gtrsim 0.1$).  We note that same behaviour has
 been also detected in the two leptohadronic models.
A study of the fitting parameters reveals that for
protons producing the
TeV $\gamma-$rays the condition $t_{\rm cool} \gg t_{\rm esc}=\tcr$ holds,
thus the system responds fast to the imposed variations
due to the small value of the
proton escape time.
On the other hand, electrons radiating in the
X-rays have a much faster cooling than escape
timescale. However,
their PSD  produces also a  break at about the same frequency
as in the TeV $\gamma-$rays case. This behaviour seems puzzling at first;
one would expect that if the cooling of particles is fast, then the photon lightcurve (in our
case the X-ray one)
would track in detail the source variability. Moreover, one would expect to find
a break in the PSD spectrum only for the TeV $\gamma$-rays, which is not the case.
In fact, it is not the cooling timescale alone that determines
the variability pattern of photons, but the minimum timescale between $t_{\rm esc}$
and $t_{\rm cool}$. As is the case for all one-zone models, the fastest possible
variation of the system is controlled by the photon crossing
timescale $\tcr$. When min$(t_{\rm esc},t_{\rm cool}) \approx \tcr$, as in our simulations, 
the photon lightcurves follow, in general, the source variations. However, the small timescale
variations are smoothed out. Even in the extreme case of ultra fast cooling ($t_{\rm cool}  << \tcr$)
photons cannot attain the full amplitude of particle
variations -- see Appendix B, and this results in a breaking
 frequency (which corresponds to a few $\tcr$) in the PSD. 
Another parameter that affects the shape of the output PSD is the amplitude
of the imposed variations -- see parameter $f$ in eq. (\ref{var}). 
For some very small value of $f$, which in our simulations is $f=0.001$, we find that the leptohadronic
system can follow the variations at injection, i.e. the PSD of the light curves
has the same shape, apart from a normalization factor, with the one of injection. However, for so small f 
values the variability is at a very low level,
and therefore for all practical purposes the source can be
considered as not varying.
% the proton injection compactness varies at each $\tcr$ which is
% the fastest timescale of the system. 
% If we, however, repeat the same simulation but with an injection varying in a larger timescale, 
% e.g. 30 $\tcr$, then we find that 
% the break of the PSD moves to the left ($f_{\rm br} \approx 1/30$).
%% Figure 2 in Dec051212 schema 

\begin{figure}
 \centering
\resizebox{\hsize}{!}{\includegraphics{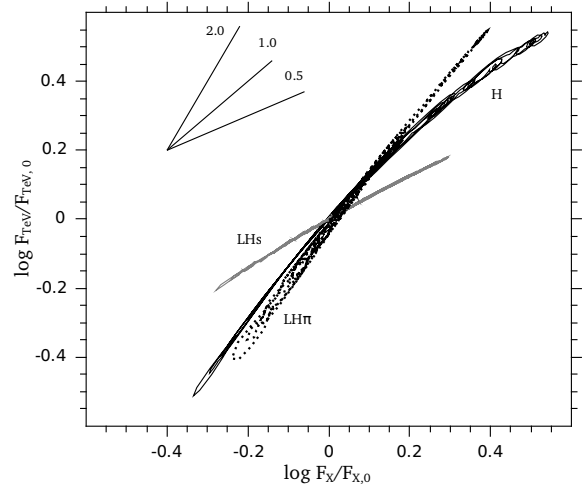}}
\caption{Plot of the TeV vs. X-ray fluxes obtained after varying only $\lpinj$ (Model H) and
both $\lpinj$ and $\leinj$ (Models LH$\pi$ $\&$ LHs). The fluxes are normalized with respect to their values of the 
pre-flaring state fit. Segments with different slopes are also plotted for reference.}
\label{fxfg-all3}
\end{figure}

For the two leptohadronic cases (Models LH$\pi$ and LHs in Fig.~\ref{MW-all}) we have implemented the
same variation\footnote{Not only the type of variation but also the series of random
numbers used in all cases is the same, unless stated otherwise.} for $\lpinj$, which was further extended to $\leinj$, since 
two types of particles are being injected in the source. These simulations generated light curves similar to those shown
in Fig.~\ref{lc-hadronic}. 
More information, however, can be deduced from flux-flux diagrams, as the one shown in Fig.~\ref{fxfg-all3} where 
the TeV flux is plotted against the 
X-ray flux for all the cases discussed so far. 
First let us focus on model H (black solid line).
%Now let us focus on the relation between the X-and $\gamma$-ray light curves.
While the X-rays follow linearly the proton synchrotron radiation,
and therefore the proton luminosity,
the TeV emission shows an almost quadratic dependence on $\lpinj$ 
for the reasons explained earlier -- see also DMPR.
% This dependence is better depicted in Fig.~\ref{fxfg-all3},
% where the TeV flux 
% is plotted against the 
% X-ray flux for all the cases discussed so far.
% Let us first focus on model H (black solid line). 
Although at low fluxes one finds a clear quadratic dependence, this
becomes flatter at higher fluxes due to the increasing
effect of $\gamma\gamma$ absorption which gradually depletes 
the TeV regime. 
Another feature that is worth mentioning,
 is the tight correlation 
between the fluxes, which is derived with no requirements
of fine tuning, same as in the SSC scenario.

% Although these simulations generated their own lightcurves, they do not 
% differ qualitatively from the pure hadronic case. Thus, it is far more instructive to look 
% directly at the TeV vs. X-ray flux diagrams for each of the three cases (see Fig.~\ref{fxfg-all3}). 
In general, 
the same trend holds also for the LH$\pi$ model (dashed line), while the LHs case
(grey solid line) leads to a (sub)linear correlation\footnote{A linear regression fit to the
data for the LHs model gives a slope  $\sim 0.7$; we characterize this
as a sublinear correlation. For a particular model, the exact slope of the $F_{\rm TeV}/F_{\rm X}$ 
correlation depends also on the energy bands considered.
For example, for the LHs model we derive an exactly linear 
correlation between the peak fluxes of the two 
distinct emission components.}. 
In the leptohadronic
models we find no break in the TeV/X-ray correlations,
since the effects of $\gamma \gamma$ absorption are less significant. 
This comes from the fact that, for both models, 
we have obtained good fits using much smaller $\lpinj$
and higher Doppler factors than in the pure hadronic one -- see Table~\ref{table1}.
Finally, in all three models, no lags between the X-ray and TeV photons were detected.

%% Figure 3 in Dec051212 schema - placeholder

\begin{figure}
 \centering
 \begin{tabular}{l}
\includegraphics[width=7.5cm, height=6.2cm]{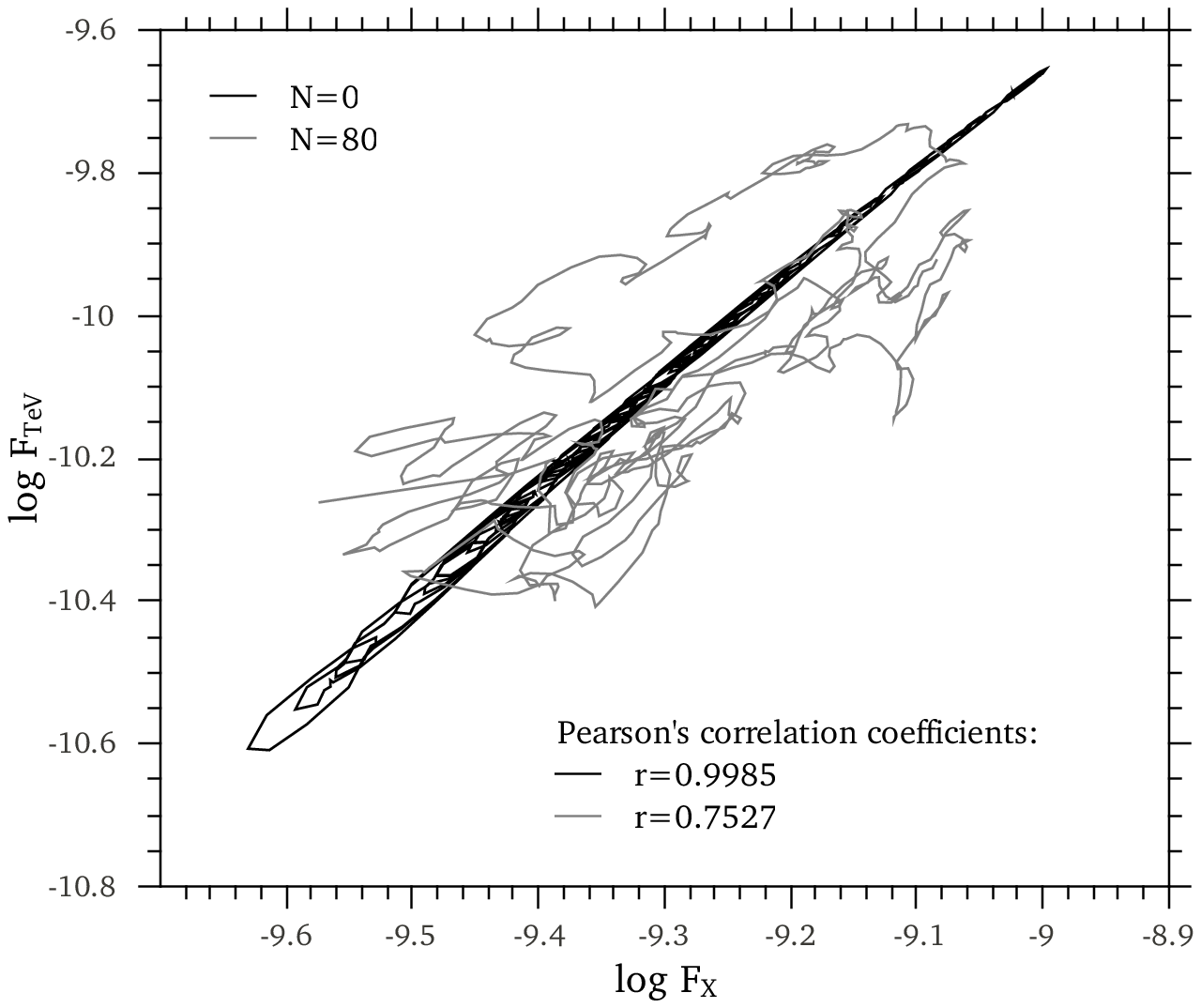} \\
 \includegraphics[width=7.cm, height=6.cm]{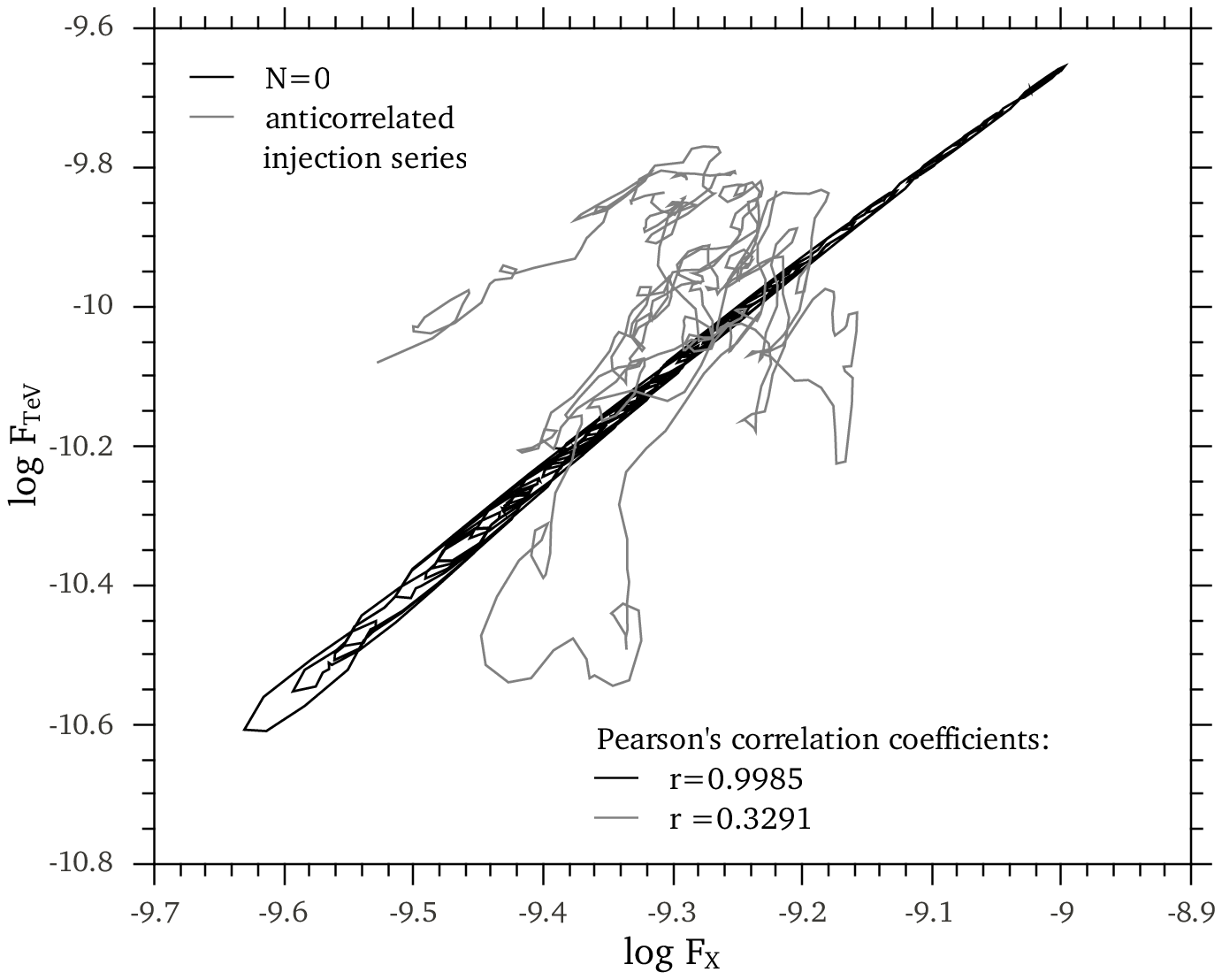} 
  \end{tabular}
%\resizebox{\hsize}{!}{\includegraphics{./figs/fig4.eps}}
\caption{Top panel: Plot of the TeV vs. X-ray fluxes obtained within 
the LH$\pi$ model, after varying $\lpinj$ and $\leinj$
with $N=0$ (black line) and $N=80$ (grey line). 
Bottom panel: Same as above except for  
the grey line, which is obtained 
using as input for $\lpinj$ and $\leinj$ two different anticorrelated time-series.
Inset legends show the Pearson's correlation coefficients for each case.}
\label{pionN0N80}
\end{figure}

\begin{figure}
 \centering
\includegraphics[width=7.5cm, height=6.2cm]{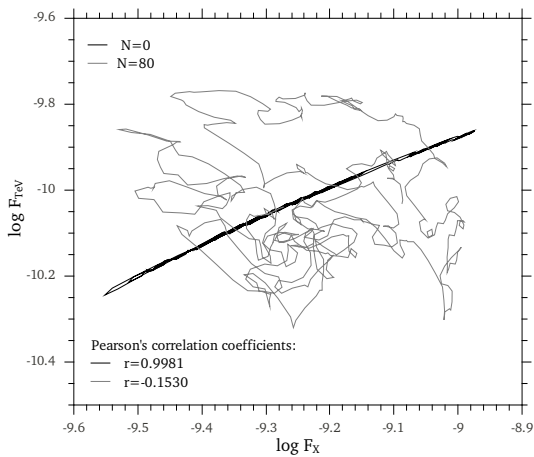} 
\caption{Plot of the TeV vs. X-ray fluxes obtained within the LHs model, 
after varying $\lpinj$ and $\leinj$ with $N=0$ (black line) and $N=80$ (grey line). Inset legend
same as in Fig.~\ref{pionN0N80}.}
\label{psynN0N80}
\end{figure}

Previous analyses of Mrk 421 have noted the presence of lags in its spectral evolution 
(e.g. \cite{Takahashi2000, Fossati2008, Singh2012}). 
Such lags cannot be reproduced by one-particle population models, such as the purely hadronic one. 
However, in leptohadronic models where two primary particle populations 
are being injected into the source, these lags can be simulated 
by introducing a shift ${\rm N}$ in the temporal variation of $\leinj$ compared to $\lpinj$,
i.e. $(\leinj)_{\rm i}=(\lpinj)_{\rm i+N}$, where the subscript denotes that quantities are calculated
at time $t_{\rm i}$ (in units of $\tcr$).

Figure \ref{pionN0N80} depicts the TeV vs. X-ray flux obtained in the LH$\pi$ case. In both panels,
the black line corresponds to $N=0$, i.e. $\lpinj$ and $\leinj$
are completely correlated. 
The grey line in the top panel is the result of a simulation, where
$\leinj$ was obtained by a relative shift of 80 $\tcr$
with respect to $\lpinj$. In this way we imposed less correlated variations
on particle injection, having a Pearson's correlation coefficient $r=0.63$.
%here, $\leinj$ was obtained by a relative shift of 80 $\tcr$
%with respect to $\lpinj$. 
We have used a variety of positive correlated $\lpinj,\leinj$ by introducing
various shifts, e.g $N=312, 512, 1024$. 
In all cases, which we do not present here, we derived large correlation coefficients
($r\gtrsim 0.5$) which show that the TeV and X-ray fluxes 
retain their strong correlation.
Then, we considered a more extreme case, where the proton and electron injection have
strong anticorrelation. For this, 
we have used two random number series with correlation coefficient $r=-0.7$. 
The resulting TeV/X-ray correlation
is shown with grey line in the bottom panel. The respective diagram
for the LHs model and $N=80$ are shown in Fig.~\ref{psynN0N80}.
% Figure \ref{pionN0N80} depicts the TeV vs. X-ray flux obtained in LH$\pi$ (top panel)
% and LHs (bottom panel) models, for $N=0$ (black line) and $N=80$ (grey line). 

In general, the introduction of a shift loosens the $F_{\rm TeV} - F_{\rm X}$ correlation. 
The effect, however,
is more evident in the proton synchrotron case where
the TeV/X-ray flux correlation is almost destroyed even for $N=80$ ($r=-0.15$).
On the other hand, in the LH$\pi$ model, even if the input series are anticorrelated, we
find a larger absolute value for the correlation coefficient -- see inset legend of Fig.~\ref{pionN0N80}.
In other words, if the 
$\gamma$-rays are modelled by the emission of secondaries 
produced by photohadronic processes and therefore any variations of
$F_{\rm TeV}$ reflect only indirectly the changes in the proton injection
rate, the correlation between the fluxes in the X-rays and TeV $\gamma$-rays
is partially retained. On the other hand, in the LHs model the variability
observed in both X-rays and $\gamma$-rays reflects directly the variability pattern of 
the particle injection rate. Thus, if there is a degree of decorrelation in the injection it will be seen
also in the flux-flux diagram.
%\footnote{We note that the strength of the $F_{\rm TeV}/F_{\rm X}$ correlation 
%depends also on our assumption of fast escape, i.e. on $\tpesc=\teesc=\tcr$. If we have considered
%slower particle escape, cooling effects would have also played a role.}.
For the LH$\pi$ case (top panel) the introduction of a shift 
decreases the maximum/minimum flux values in both energy bands, since the 
$\gamma$-ray luminosity depends also on the number density of soft
target photons, e.g. synchrotron photons in the X-rays. 
In the LHs case on the other hand, 
where the $\gamma$-rays are the product of proton synchrotron radiation, 
the $\gamma$-ray
flux does not depend on the X-ray photons, as long as the
X-ray photon number density is low enough as not to cause
significant $\gamma \gamma$ absorption. 
 Thus, the
range of flux variations remains approximately the same.
% Large shifts in the time series of $\leinj$ with respect to $\lpinj$ are able 
% to destroy any correlation, even in the photopion case, provided that the two variations do not happen to enter back in phase by chance.

\subsection{Varying $\gamma_{\rm max}$}
\begin{figure}
 \centering
\resizebox{\hsize}{!}{\includegraphics{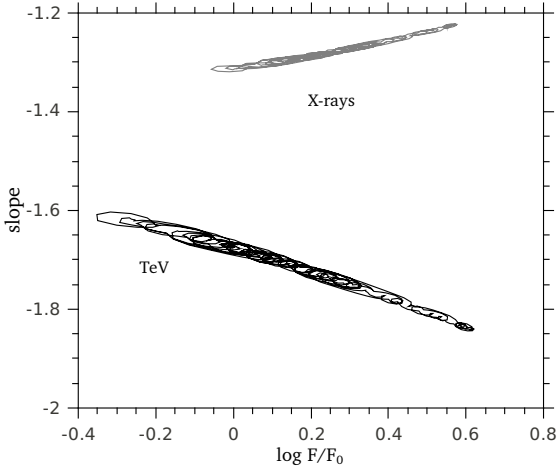}}
\caption{Plot of the spectral index $\beta$ as a function of the flux 
in the X-rays (grey line) and in the TeV energy regime (black line) for the photopion
model, in the case where the injection compactness is the varying
parameter. The fluxes are normalized with respect to their values obtained
for the pre-flaring fit shown in Fig.~\ref{MW-all}. }
\label{fig6}
\end{figure}

Observations of Mrk 421 during different periods
of flaring activity indicate spectral evolution, and
in particular spectral hardening in the X-rays and/or in the TeV $\gamma$-rays
(e.g. \citealt{Takahashi2000, Fossati2008}). 
Variations of the
injection compactness alone cannot reproduce these observational
findings. This is  exemplified in Fig.~\ref{fig6}, where 
the spectral indices\footnote{We define the spectral index 
as $F_{\nu} \propto \nu^{\beta}$.} in the X-ray (grey line) and TeV energy bands (black line),
are plotted against the corresponding fluxes, for one of our fitting models (LH$\pi$).
The X-ray spectral index remains approximately constant during flux variations, while
% Although we find a slight hardening of the X-ray spectrum as the flux increases, this cannot 
% be resolved by any measurement so far. Thus, within the errorbars of typical observations
% no spectral evolution in the X-rays would be observed. 
% Moreover, we find that under variations of the injection compactness
the $\gamma$-ray part of the spectrum becomes softer during flaring events. This is to be expected,
since the TeV observations were fitted by the cutoff of the synchrotron component, in  all three of our
models. We note also that if $\gamma \gamma$ absorption is significant in the TeV regime,
one expects to find an almost constant spectral index, even though the injection compactness
varies. These two  features, i.e. absence of spectral evolution in the X-rays and/or spectral softening
in the $\gamma$-ray regime, are also obtained for the pure hadronic and the proton synchrotron case.

\begin{figure}
 \centering
\resizebox{\hsize}{!}{\includegraphics{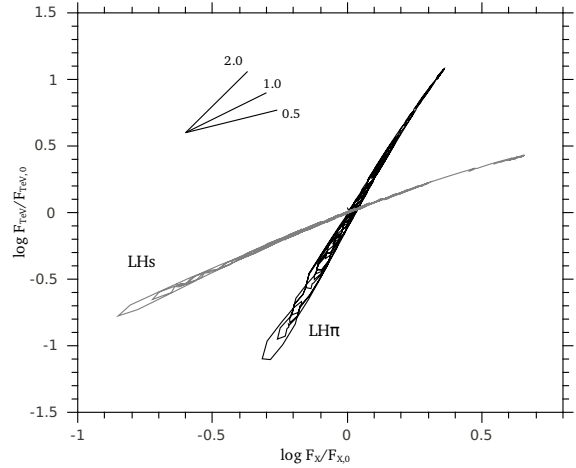}}
\caption{Plot of the TeV vs. X-ray fluxes obtained after varying both $\gpmx$ and $\gemx$
 in Models LH$\pi$ $\&$ LHs shown in black and grey color respectively. 
The fluxes are normalized with respect to their values of the 
pre-flaring state fit. Segments with different slopes are also plotted for reference.}
\label{flux-flux-gmx}
\end{figure}

\begin{figure}
  \centering
 \begin{tabular}{c}
\includegraphics[width=8.cm, height=6.cm]{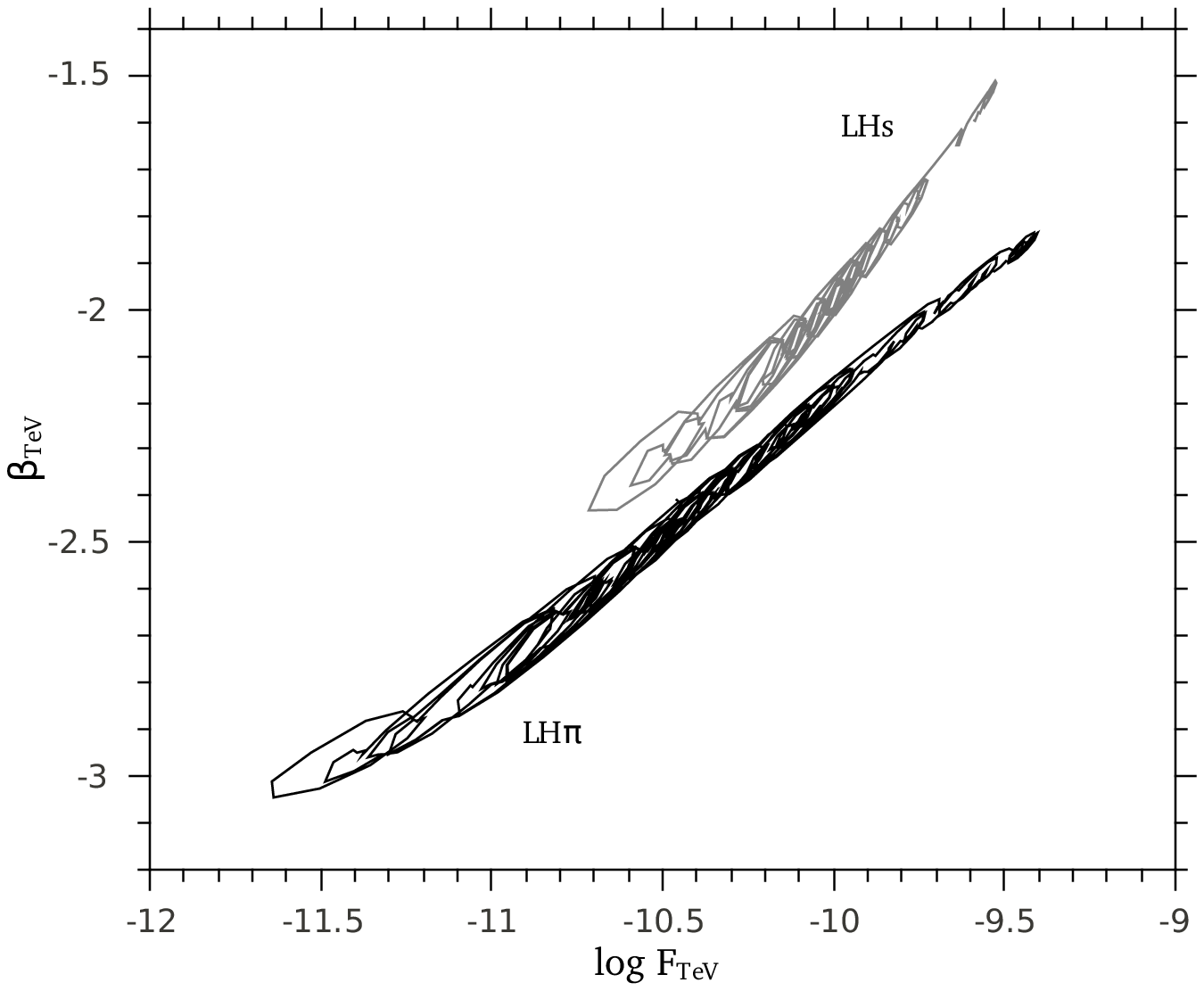} \\
\includegraphics[width=7.5cm, height=6.cm]{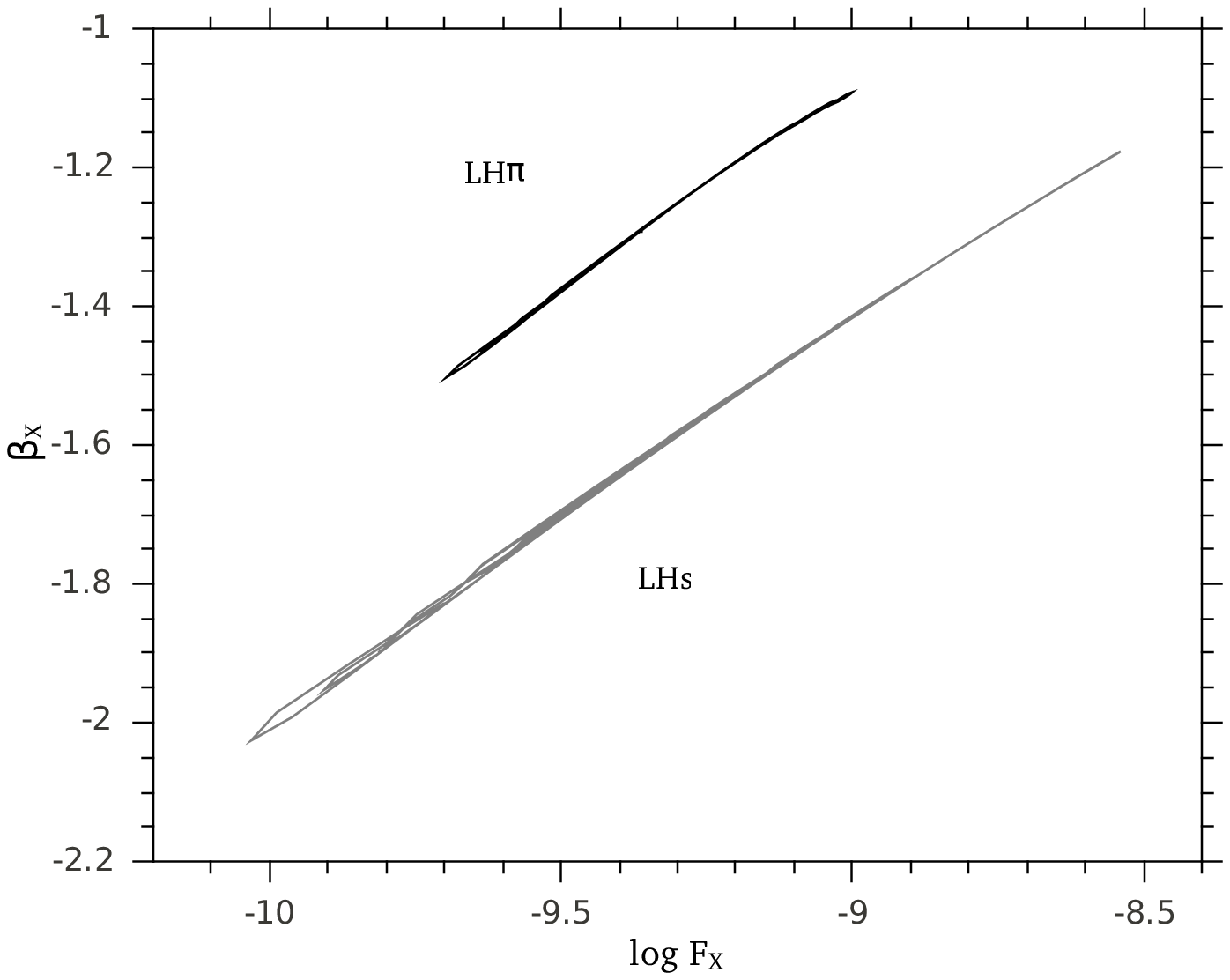} 
  \end{tabular}
\caption{Plot of the spectral index $\beta$ as a function of the flux $F$
in the TeV energy band (top panel) and in the X-rays (bottom panel). 
Black and grey lines correspond to the 
LH$\pi$ and LHs case respectively.
For clarity reasons we have shifted the grey curve in the top panel
by $0.2$ units upwards.}
\label{slope-flux}
\end{figure}

The second varying parameter that we have studied  is
the maximum energy of protons and electrons. We have kept
constant the injection compactnesses and we have once
again adopted the same random number series as in \S 3.2 for
both $\gemx$ and $\gpmx$.
In Figs. \ref{flux-flux-gmx}  and 
\ref{slope-flux} we show indicatively the
results for the leptohadronic cases. 
For the proton synchrotron case, the linear flux-flux correlation is retained, in 
contrast to the photopion case. For the latter, we find a correlation 
 between the TeV and X-ray fluxes, 
which is steeper than the quadratic we have obtained by varying $\leinj$ and $\lpinj$. 
Actually, a fit to our results shows that $F_{\rm TeV} \propto F_{\rm X}^{3.3}$.
In the LH$\pi$ model the exact slope of the TeV/X-ray flux correlation is 
sensitive to the power-law exponent of the electron distribution. If the power-law spectrum 
of electrons at injection is flat, as in our LH$\pi$ fit (see Table \ref{table1}), any variations
of $\gemx$ result in non-negligible variations of the injection rate, since $\leinj$ is kept constant.
Thus, the number density of synchrotron photons emitted by electrons with $\gamma <\gemx$, which
serve as targets for protons, vary significantly. In such case, we expect that the $\gamma$-ray emission
will show also large variations, since it depends on both the soft photon field
and the proton distribution. On the other hand, 
when fitting the SED with a steeper power-law electron distribution (e.g. $p_{\rm e}=p_{\rm p}=1.5$),
the injection rate of lower energy electrons remains approximately constant. In this case,
we obtain a quadratic relation between TeV and X-ray fluxes. 
%\textbf{WHY?}.

Following the same procedure as in \S 3.2, we have then studied the effects of partially correlated 
variations of $\gpmx$ and $\gemx$
on the TeV/X-ray correlation. We find that the degree of TeV/X-ray flux correlation
is more easily destroyed in the LHs model. As for the LH$\pi$ model is concerned, we find, 
for the same shift, a more tight correlation in the case of variable injection than of variable energy cutoff.

The TeV/X-ray correlations obtained so far (for correlated input time-series)
are summarized in Table \ref{table2}. The symbol `$\sim$' is used
to denote possible deviations from the exact
trend due to effects that do not 
depend on the fitting model itself, such as
$\gamma \gamma$ absorption and the 
particular choice of the energy bands.

\begin{table}
\centering
\caption{Slope of the TeV/X-ray correlations obtained 
for variations on two model parameters.}
\begin{tabular}{c ccc}
\hline
%\phantom{} &   Model 1 & Model 2 & Model 3 \\
Parameter & Model H & Model LH$\pi$ & Model LHs \\
\hline \hline
$\ell_{\rm inj}$ & linear to&  quadratic & $\sim$  linear \\
\phantom{} & quadratic  &  \phantom{} & \phantom{} \\
\hline
 $\gamma_{\rm max}$ & quadratic & quadratic to & $\sim$ linear\\
\phantom{} & \phantom{} & cubic  & \phantom{}\\
\hline
\end{tabular}
\label{table2}
\end{table}

\begin{figure}
\centering
\resizebox{\hsize}{!}{\includegraphics{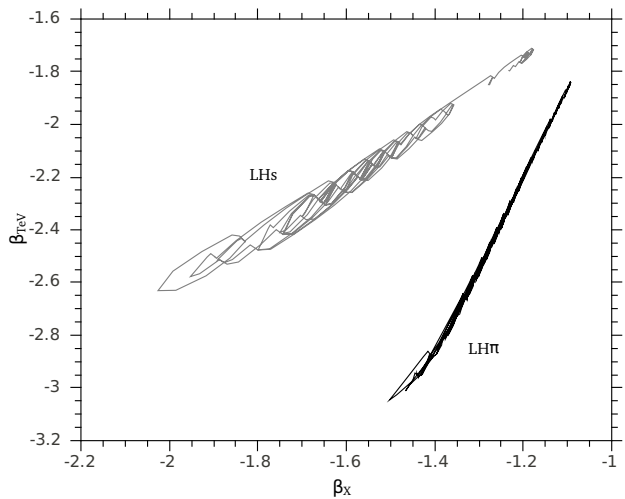}}
\caption{Plot of the spectral index in the TeV energy range ($\beta_{\rm TeV}$)
vs. the corresponding one in the X-rays ($\beta_{\rm X}$) for the 
LH$\pi$ (black line) and LHs (grey line) models.}
\label{btev-bx}
\end{figure}

Figure \ref{slope-flux} shows that an increasing flux in both
X-rays (bottom panel) and $\gamma$-rays (top panel) is accompanied by a hardening of the spectrum;
this trend is in good agreement with
many observations regarding spectral variability of Mrk 421.
Figure \ref{slope-flux} reveals also a `mirror' symmetry between the photopion and proton synchrotron cases, i.e.
the range of flux and spectral variations is larger in the TeV energy band for the LH$\pi$ model 
and in the X-rays for the LHs. 
More specifically, the fact that $\beta_X$ varies
less in the photopion case than in the proton synchrotron one can be used as
diagnostic tool between the two models. 
The range of spectral variations in the
two models is better displayed in Fig.~\ref{btev-bx}, where
the spectral index in the TeV energy range is plotted
against the one in the X-rays. Although $\beta_{\rm TeV}$ varies approximately
between the same values in both models, spectral variations 
in the X-rays are more prominent in the proton synchrotron model.
In particular, we find  $\beta_{\rm TeV} \propto 3.3 \beta_{\rm X}$ and
$\beta_{\rm TeV} \propto 1.1 \beta_{\rm X}$ for LH$\pi$ and LHs 
respectively.

The X-ray spectrum in the photopion case is relatively hard and shows smaller spectral variations
due to the Bethe-Heitler component. If we artificially switch-off the channel of photopair
production, find a fit to the pre-flaring state of 
Mrk 421 and then produce variations to $\gemx/\gpmx$ as before, we find that
the range of variations for both  $\beta_{\rm X}$ and 
$F_{\rm X}$ increases; 
specifically, $\beta_{\rm X}$ and $\log F_{\rm X}$ varry between $(-1.7, -1)$ and $(-9.9, -8.5)$ respectively, while their
correlation law remains unaffected.
Therefore, if the emission from the Bethe-Heitler process is neglected, 
 no significant discrepancy between  LH$\pi$ and LHs models is predicted.

\section{Summary/Discussion}
In the present paper we have examined the spectral and variability signatures of the
so-called (lepto)hardronic models. These models are routinely used to model
MW observations of high energy blazars and constitute a viable alternative
to the leptonic ones. According to standard practice, the low frequency
part of the spectrum (usually up to the X-ray regime) is fitted by electron
synchrotron radiation, while the higher part is fitted by hadronic emission -- most
commonly proton synchrotron radiation.
 
For the purposes of the present treatment we have used the 
time-dependent numerical code
presented in DMPR. This models the photopair and photopion processes in
great detail by using results from the Monte Carlo codes of  
\cite{protheroejohnson96} and \cite{SOPHIA2000}
respectively for the production rates of secondaries.
We have chosen, as an illustrative example, 
to fit the well monitored TeV blazar Mrk 421 during a flaring state
using the results of the 2001 campaign \citep{Fossati2008}. 
For this, we followed the usual algorithm (for a similar approach
to the leptonic case, see \cite{mastkirk97,  krawczynskietal02})
where we first fitted the SED for
a pre-flaring
state and then we simulated variability by varying some key parameter
of the fit -- in our case it was either the particle injection
luminosity or the upper cutoff of the particle distribution.

As explained in DMPR, hadronic models have three important processes
which can lead to high energy photon emission. These are  proton synchrotron
radiation, photopair, and photopion production.     
In principle all three could be used, in various combinations, in attempting a fit to 
the SED of a $\gamma-$ray blazar.
In practice we found that there are three combinations which 
give satisfactory results to the Mrk 421 case: 
\begin{enumerate}
\item The `pure hadronic' (H) model:
In this case
there is no need for a leptonic component.
The X-rays are produced
from proton synchrotron radiation while the TeV $\gamma$-rays 
were `pion induced' -- with this we mean that $\gamma$-rays
are produced from a combination of the
synchrotron radiation of electrons produced in charged pion decay
and of the electromagnetic cascade induced from neutral pion decay
$\gamma-$rays.  
\item The 'leptohadronic pion' (LH$\pi$) model: 
The X-rays are produced from the synchrotron 
radiation of a primary leptonic
component  while the $\gamma$-rays are once again pion-induced.
\item The 'leptohadronic synchrotron' (LHs) model: 
Here the X-rays are produced as in the previous model
while the $\gamma$-rays are produced by proton synchrotron radiation.
\end{enumerate} 
All three models need high magnetic fields -- as compared to the pure
leptonic fits, and high values of $\gpmx$ with the LHs model requiring
the most extreme values of both parameters. However, this is also
the model that by far is the more `economic' for the required energy
density in relativistic protons -- see Table \ref{table1}.

As a next step,  we introduced random-walk type perturbations either in
the compactness of the injected particles  or in their upper cutoff.
We have assumed that each consecutive value of these parameters is
either increased or decreased by 5\% over its previous value. 
If many of these small amplitude 
perturbations pile up they can produce a `statistical' flare
and we have chosen such an example to test the variability
signatures of the models listed above
(see Fig.~\ref{lc-hadronic})  -- note however that nowhere  does the
varying parameter exceed its initial value by more than a factor of 2.
Feeding the values of the varying parameter
to the code we produce simulated variations
in the X-ray and TeV $\gamma$-ray bands and searched for
possible correlations between their respective fluxes
and spectral indices.

In the case of a varying injection compactness, the H model gives us
a clear quadratic dependence between the X-ray and TeV $\gamma$-rays
-- this can be easily explained from the fact that the H model
resembles in many aspects to the SSC leptonic one (see DMPR).
An interesting
feature is that at high fluxes $\gamma\gamma$ absorption tends to
linearize the correlation -- see Fig.~\ref{fxfg-all3}.
The other two leptohadronic models need at least two more free parameters
to fully specify their variability. One relates the amplitudes of variation
between protons and electrons and the other their phases. Here we have
examined the most simple cases, i.e. we have assumed equal amplitude 
changes between the two species which can be either contemporaneous
or with a time shift of several $\tcr$.
In  the no-lag case, the LH$\pi$
model produces a quadratic dependence between the fluxes,
while the LHs a less than linear one. Note that because of the
values of the fitting parameters (see Table \ref{table1}) $\gamma \gamma$
absorption is more severe in the latter case than in the former,
a fact that explains the curvature seen in the model LHs curve.
In the case where we have allowed time lags, then both 
models loosen up their correlation. We found however that in the
case of the LH$\pi$ model a general quadratic trend remains while for
the LHs model all correlation is lost. Finally, all three models
seem to get steeper TeV spectra as the flux increases -- see Fig.\ref{fig6}.

In the case where the maximum energy of the particle distribution was
varying, we found that the results 
are similar to the particle injection case, with the notable difference that all models produce
a spectral hardening to both X-rays and TeV $\gamma-$rays. We have
also found that time shifts tend to decorrelate more the lightcurves,
however the effect is stronger in the LHs model than in the LH$\pi$.

Obtaining the PSD of the light curves we find that they resemble
the one of injection, up to a break frequency that corresponds to a few 
$\tcr$. In particular, for both TeV and X-ray light curves we find less power at the high-frequency
part of their PSD (small timescale variations) when compared to that of the source.
In other words, for the parameters used in the present work (see Table \ref{table1}), the photon field
cannot react faster that a few $\tcr$ to the imposed variations -- see section 3.2 and Appendix B for more
details. Note also that the break frequency of the PSD corresponds to $\sim 0.5-1$ days,
 which
does not contradict other published results (e.g. \cite{Takahashi2000, kataokaetal01}).

Although we have not addressed in detail the acceleration
mechanism, the adopted modelling for the variations corresponds
to a physical picture, where particles first are being
accelerated close to a shock front (this region is a `black box' in our model) and then, they
are being injected in the emitting volume. Thus, any changes that occur
in the region where acceleration takes place, are later seen as changes in the
characteristics of the particle injection mechanism. These on their
turn lead to an outbursting behaviour of the source. Studying
the characteristics of this flaring activity was the actual aim of the
present work. In this respect, the introduced time-lags 
play the role of some intrinsic difference in the 
acceleration mechanism between the two species which, however, retains
some coherence.
Two-zone models which have been applied to 
leptonic models  (\cite{kirima98} and \cite{moraitismast11})
are more elaborate as far as
particle acceleration is concerned, however they can become very 
cumbersome when treating hadronic processes.

Concluding we can say that
it is very interesting that all three models give very good $\chi^2$ fits
to the Mrk 421 data --
the two leptohadronic models have slightly better fits, however one should
have in mind that they use
many more free parameters.  We note also that it is the first time that
the pion-induced $\gamma$-ray component is used in fitting TeV data. 
In this case we found that a broad feature in the low to 
medium $\gamma-$rays is produced from the synchrotron radiation 
of  Bethe-Heitler pairs (see Fig. 1).
This feature
is very restrictive as far as fitting is concerned, but its presence
might provide a decisive observational clue for the viability of
this brand of hadronic models as it is not present neither in
the LHs nor in the leptonic SSC models. 
We also found that proton synchrotron models are favoured as far as energetics
is concerned. However, we note that the H and LH$\pi$ models
produce in a much more natural way correlated variability
between X-rays and TeV $\gamma-$rays which is mostly quadratic
-- that is they require equal amplitude variations between the particles
and can allow for shifts in their phases. 
Proton synchrotron
models, on the other hand, are inherently linear and therefore, in order 
to produce quadratic correlations they need the rather special conditions
that the electron injection varies quadratically with respect to the
one of the protons. Furthermore,
they require  a tight phase correlation between the injection of
the two species
as all coherence is lost even in short time shifts. 

The analysis presented here deals with variability signatures in the framework of leptohadronic models, such as 
TeV/X-ray correlation, spectral evolution in the X-rays and very high energy (VHE) $\gamma$-rays and
PSD of the corresponding light curves. These could be used as a diagnostic tool for 
differentiating between the models and therefore probing the nature of high-energy radiating particles.
Observational requirements for this are the high temporal and spectral resolution in the VHE part of the $\gamma$-ray spectrum along with
contemporaneous X-ray observations. Great importance for the first requirement will be 
the future Cherenkov Telescope Array (CTA), since it will have high sensitivity and will provide us with 
large count rate light curves on a routine daily basis rather than on exceptional flaring events only 
\citep{Sol2012}.
% The analysis presented here deals exclusively with hadronic systems
% in the sub-critical regime. Preliminary calculations show that
% if the imposed variations are larger, then they can drive the
% system to supercriticality. This will result in the appearance
% of highly non-linear outbursts in the photon lightcurves
% which will create QPO-like features
% in the PSD as the supercriticality tends to amplify certain modes.
% We will deal with this in a forthcoming publication.

\section{Acknowledgments}
We would like to thank Dr. D. Emmanoulopoulos for useful discussions
and comments on the manuscript.
This research has been co-financed by the European Union (European Social Fund – ESF) and Greek national funds through the Operational 
Program ``Education and Lifelong Learning'' of the National Strategic Reference Framework (NSRF) - 
Research Funding Program: Heracleitus II. 
Investing in knowledge society through the European Social Fund.

\bibliographystyle{mn2e} % style mn2e.bst
\bibliography{mpd2012} % your references Yourfile.bib

\begin{thebibliography}{35}
\expandafter\ifx\csname natexlab\endcsname\relax\def\natexlab#1{#1}\fi

\bibitem[{{Abdo} {et~al}\mbox{.}(2011){Abdo}, {Ackermann}, {Ajello}, {Baldini},
  {Ballet}, {Barbiellini}, {Bastieri}, {Bechtol}, {Bellazzini}, {Berenji}, \&
  et~al.}]{abdoetal11}
{Abdo} A.~A. {et~al.}, 2011, ApJ, 736, 131

\bibitem[{{Aharonian}(2000)}]{aharonian00}
{Aharonian} F.~A., 2000, New Astron., 5, 377

\bibitem[{{Boettcher}(2012)}]{boettcher12}
{Boettcher} M., 2012, ArXiv e-prints

\bibitem[{{Boettcher} \& {Dermer}(1998)}]{boettcherdermer98}
{Boettcher} M., {Dermer} C.~D., 1998, ApJ, 501, L51

\bibitem[{{Dermer} \& {Schlickeiser}(1993)}]{dermerschlickeiser93}
{Dermer} C.~D., {Schlickeiser} R., 1993, ApJ, 416, 458

\bibitem[{{Dermer} {et~al}\mbox{.}(1992){Dermer}, {Schlickeiser}, \&
  {Mastichiadis}}]{dermeretal92}
{Dermer} C.~D., {Schlickeiser} R., {Mastichiadis} A., 1992, A\&A, 256, L27

\bibitem[{{Dimitrakoudis} {et~al}\mbox{.}(2012){Dimitrakoudis}, {Mastichiadis},
  {Protheroe}, \& {Reimer}}]{DMPR2012}
{Dimitrakoudis} S., {Mastichiadis} A., {Protheroe} R.~J., {Reimer} A., 2012,
  A\&A, 546, A120

\bibitem[{{Emmanoulopoulos} {et~al}\mbox{.}(2010){Emmanoulopoulos}, {McHardy},
  \& {Uttley}}]{emmanoulopoulosetal10}
{Emmanoulopoulos} D., {McHardy} I.~M., {Uttley} P., 2010, MNRAS, 404, 931

\bibitem[{{Fossati} {et~al}\mbox{.}(2008){Fossati}, {Buckley}, {Bond},
  {Bradbury}, {Carter-Lewis}, {Chow}, {Cui}, {Falcone}, {Finley}, {Gaidos},
  {Grube}, {Holder}, {Horan}, {Horns}, {Jordan}, {Kieda}, {Kildea},
  {Krawczynski}, {Krennrich}, {Lang}, {LeBohec}, {Lee}, {Moriarty}, {Ong},
  {Petry}, {Quinn}, {Sembroski}, {Wakely}, \& {Weekes}}]{Fossati2008}
{Fossati} G. {et~al.}, 2008, ApJ, 677, 906

\bibitem[{{Fossati} {et~al}\mbox{.}(1998){Fossati}, {Maraschi}, {Celotti},
  {Comastri}, \& {Ghisellini}}]{fossatietal98}
{Fossati} G., {Maraschi} L., {Celotti} A., {Comastri} A., {Ghisellini} G.,
  1998, MNRAS, 299, 433

\bibitem[{{Ghisellini} \& {Madau}(1996)}]{ghisellinimadau96}
{Ghisellini} G., {Madau} P., 1996, MNRAS, 280, 67

\bibitem[{{Kataoka} {et~al}\mbox{.}(2001){Kataoka}, {Takahashi}, {Wagner},
  {Iyomoto}, {Edwards}, {Hayashida}, {Inoue}, {Madejski}, {Takahara},
  {Tanihata}, \& {Kawai}}]{kataokaetal01}
{Kataoka} J. {et~al.}, 2001, ApJ, 560, 659

\bibitem[{{Kirk} \& {Mastichiadis}(1992)}]{kirkmasti92}
{Kirk} J.~G., {Mastichiadis} A., 1992, Nature, 360, 135

\bibitem[{{Kirk} {et~al}\mbox{.}(1998){Kirk}, {Rieger}, \&
  {Mastichiadis}}]{kirima98}
{Kirk} J.~G., {Rieger} F.~M., {Mastichiadis} A., 1998, A\&A, 333, 452

\bibitem[{{Konopelko} {et~al}\mbox{.}(2003){Konopelko}, {Mastichiadis}, {Kirk},
  {de Jager}, \& {Stecker}}]{konopelkoetal03}
{Konopelko} A., {Mastichiadis} A., {Kirk} J., {de Jager} O.~C., {Stecker}
  F.~W., 2003, ApJ, 597, 851

\bibitem[{{Krawczynski} {et~al}\mbox{.}(2002){Krawczynski}, {Coppi}, \&
  {Aharonian}}]{krawczynskietal02}
{Krawczynski} H., {Coppi} P.~S., {Aharonian} F., 2002, MNRAS, 336, 721

\bibitem[{{Mannheim} \& {Biermann}(1992)}]{mannheimbiermann92}
{Mannheim} K., {Biermann} P.~L., 1992, A\&A, 253, L21

\bibitem[{{Maraschi} {et~al}\mbox{.}(1992){Maraschi}, {Ghisellini}, \&
  {Celotti}}]{maraschietal92}
{Maraschi} L., {Ghisellini} G., {Celotti} A., 1992, ApJ, 397, L5

\bibitem[{{Mastichiadis} \& {Kazanas}(2006)}]{mastikaza06}
{Mastichiadis} A., {Kazanas} D., 2006, ApJ, 645, 416

\bibitem[{{Mastichiadis} \& {Kazanas}(2009)}]{mastikaza09}
{Mastichiadis} A., {Kazanas} D., 2009, ApJ, 694, L54

\bibitem[{{Mastichiadis} \& {Kirk}(1997)}]{mastkirk97}
{Mastichiadis} A., {Kirk} J.~G., 1997, A\&A, 320, 19

\bibitem[{{Moraitis} \& {Mastichiadis}(2011)}]{moraitismast11}
{Moraitis} K., {Mastichiadis} A., 2011, A\&A, 525, A40

\bibitem[{{M{\"u}cke} {et~al}\mbox{.}(2000){M{\"u}cke}, {Engel}, {Rachen},
  {Protheroe}, \& {Stanev}}]{SOPHIA2000}
{M{\"u}cke} A., {Engel} R., {Rachen} J.~P., {Protheroe} R.~J., {Stanev} T.,
  2000, Computer Physics Communications, 124, 290

\bibitem[{{M{\"u}cke} {et~al}\mbox{.}(2003){M{\"u}cke}, {Protheroe}, {Engel},
  {Rachen}, \& {Stanev}}]{mueckeetal03}
{M{\"u}cke} A., {Protheroe} R.~J., {Engel} R., {Rachen} J.~P., {Stanev} T.,
  2003, Astroparticle Physics, 18, 593

\bibitem[{{Petropoulou} \& {Mastichiadis}(2011)}]{petromasti11}
{Petropoulou} M., {Mastichiadis} A., 2011, A\&A, 532, A11

\bibitem[{{Petropoulou} \& {Mastichiadis}(2012{\natexlab{a}})}]{PM12a}
{Petropoulou} M., {Mastichiadis} A., 2012{\natexlab{a}}, MNRAS, 426, 462

\bibitem[{{Petropoulou} \& {Mastichiadis}(2012{\natexlab{b}})}]{PM12b}
{Petropoulou} M., {Mastichiadis} A., 2012{\natexlab{b}}, MNRAS, 421, 2325

\bibitem[{{Protheroe} \& {Johnson}(1996)}]{protheroejohnson96}
{Protheroe} R.~J., {Johnson} P.~A., 1996, Astroparticle Physics, 4, 253

\bibitem[{{Raiteri} {et~al}\mbox{.}(2012){Raiteri}, {Villata}, {Smith},
  {Larionov}, {Acosta-Pulido}, {Aller}, {D'Ammando}, {Gurwell}, {Jorstad},
  {Joshi}, {Kurtanidze}, {L{\"a}hteenm{\"a}ki}, {Mirzaqulov}, {Agudo}, {Aller},
  {Ar{\'e}valo}, {Arkharov}, {Bach}, {Ben{\'{\i}}tez}, {Berdyugin}, {Blinov},
  {Blumenthal}, {Buemi}, {Bueno}, {Carleton}, {Carnerero}, {Carosati},
  {Casadio}, {Chen}, {Di Paola}, {Dolci}, {Efimova}, {Ehgamberdiev},
  {G{\'o}mez}, {Gonz{\'a}lez}, {Hagen-Thorn}, {Heidt}, {Hiriart}, {Holikov},
  {Konstantinova}, {Kopatskaya}, {Koptelova}, {Kurtanidze}, {Larionova},
  {Larionova}, {Le{\'o}n-Tavares}, {Leto}, {Lin}, {Lindfors}, {Marscher},
  {McHardy}, {Molina}, {Morozova}, {Mujica}, {Nikolashvili}, {Nilsson},
  {Ovcharov}, {Panwar}, {Pasanen}, {Puerto-Gimenez}, {Reinthal}, {Richter},
  {Ros}, {Sakamoto}, {Schwartz}, {Sillanp{\"a}{\"a}}, {Smith}, {Takalo},
  {Tammi}, {Taylor}, {Thum}, {Tornikoski}, {Trigilio}, {Troitsky}, {Umana},
  {Valcheva}, \& {Wehrle}}]{raiterietal12}
{Raiteri} C.~M. {et~al.}, 2012, A\&A, 545, A48

\bibitem[{{Sikora} {et~al}\mbox{.}(1994){Sikora}, {Begelman}, \&
  {Rees}}]{sikoraetal94}
{Sikora} M., {Begelman} M.~C., {Rees} M.~J., 1994, ApJ, 421, 153

\bibitem[{{Singh} {et~al}\mbox{.}(2012){Singh}, {Bhattacharyya}, {Bhatt}, \&
  {Tickoo}}]{Singh2012}
{Singh} K.~K., {Bhattacharyya} S., {Bhatt} N., {Tickoo} A.~K., 2012, New
  Astron., 17, 679

\bibitem[{Sol {et~al}\mbox{.}(2012)Sol, Zech, Boisson, de~Almeida, Biteau,
  Contreras, Giebels, Hassan, Inoue, Katarzyński, Krawczynski, Mirabal,
  Poutanen, Rieger, Totani, Benbow, Cerruti, Errando, Fallon, de~Gouveia
  Dal~Pino, Hinton, Inoue, Lenain, Neronov, Takahashi, Takami, \&
  White}]{Sol2012}
Sol H. {et~al.}, 2012, Astroparticle Physics

\bibitem[{{Stawarz} \& {Kirk}(2007)}]{stawarzkirk}
{Stawarz} {\L}., {Kirk} J.~G., 2007, ApJ, 661, L17

\bibitem[{{Takahashi} {et~al}\mbox{.}(2000){Takahashi}, {Kataoka}, {Madejski},
  {Mattox}, {Urry}, {Wagner}, {Aharonian}, {Catanese}, {Chiappetti}, {Coppi},
  {Degrange}, {Fossati}, {Kubo}, {Krawczynski}, {Makino}, {Marshall},
  {Maraschi}, {Piron}, {Remillard}, {Takahara}, {Tashiro}, {Terasranta}, \&
  {Weekes}}]{Takahashi2000}
{Takahashi} T. {et~al.}, 2000, ApJ, 542, L105

\bibitem[{{Ulrich} {et~al}\mbox{.}(1997){Ulrich}, {Maraschi}, \&
  {Urry}}]{ulrichetal97}
{Ulrich} M.-H., {Maraschi} L., {Urry} C.~M., 1997, ARA\&A, 35, 445

\end{thebibliography}
\appendix
\section[]{Latent primary leptonic component in the `H' model}
One of the models studied in the present work is the pure
hadronic (`H'), where the two main emission components
of the MW spectrum are attributed to protons.
This assumption does not exclude the 
presence of 
a primary leptonic component, as long as its contribution
to the total synchrotron emission in the X-ray regime is negligible.

Thus, let us derive an approximate upper limit 
for the injection compactness of primary
electrons $\leinj$. 
%For simplicity, we assume that all the injected power
%in primary electrons is radiated trough
%synchrotron emission, i.e. $L_{\rm e}^{\rm inj} \approx L_{\rm e}^{\rm syn}$.
The proton synchrotron radiated power can be calculated by
\eqb
L_{\rm p}^{\rm syn} = \int_{\gpmn}^{\gpmx} \textrm{d}\gamma \ n_{\rm p}(\gamma) \left.\frac{dE}{dt} \right|_{\rm p, rad},
\eqe
where 
\eqb
\left. \frac{dE}{dt} \right|_{\rm p, rad} = \frac{4}{3} \sth c \ub \left(\frac{\me}{\mpr}\right)^2\gamma^2.
\eqe
Assuming that protons at the injection have a power-law distibution, $n_{\rm p}= K_{\rm p} \gamma^{-p}$
between $\gpmn$ and $\gpmx$, with $\gpmn << \gpmx$ and $p<3$,
 the integral above results in
\eqb
L_{\rm p}^{\rm syn} \approx \frac{4}{3} \sth c \ub K_{\rm p} 
\left(\frac{\me}{\mpr}\right)^2 \frac{\gpmx^{3-p}}{3-p}.
\label{Lpsyn}
\eqe
The normalization constant $K_{\rm p}$ can be expressed in terms of the proton injection compactness -- see eq.~(\ref{lpinj})
in \S2. 
In the case were proton cooling is not significant, the proton injection luminosity is approximately given by
\eqb
L_{\rm p}^{\rm inj} & \approx & \frac{\mpr c^2 }{\tpesc} \int_{\gpmn}^{\gpmx}
 \textrm{d}\gamma \ \gamma n_{\rm p}(\gamma) \approx
\frac{\mpr c^2 }{\tcr} K_{\rm p} F_{\rm p},
\eqe
where 
\eqb
F_{\rm p} = \frac{\gpmx^{2-p}- \gpmn^{2-p}}{2-p} \approx \frac{\gpmx^{2-p}}{2-p}, \ \textrm{if} \ p<2.
\eqe

Thus, $K_{\rm p}$ is written as 
\eqb
K_{\rm p} = \frac{4 \pi R^2 \lpinj}{\sth F_{\rm p}}.
\label{Kp}
\eqe 
% 
% \eqb
% L_{\rm e}^{\rm syn} = \frac{4}{3} \sth c \ub K_{\rm e} \frac{\gemx^{3-s}}{3-s},
% \eqe
% where $s$ and $K_{\rm e}$ are the power-law index and normalization constant of the electron distribution at injection.
For the electron distribution, on the other hand, we assume that all
the injected power is radiated, i.e. $L_{\rm e}^{\rm syn} \approx L_{\rm e}^{\rm inj}$.
% 
% The latter one is given by
% \eqb
% K_{\rm e} = \frac{4 \pi R^2 \leinj}{\sth F_{\rm e}},
% \eqe
% where 
% \eqb
% F_{\rm e} = \frac{\gemx^{2-s}- \gemn^{2-s}}{2-s}.
% \eqe
% If we further assume that protons and electrons of maximum energy radiate at the X-ray regime, we find
% one relation between $\gpmx$ and $\gemx$
% \eqb
% \frac{\gpmx}{\gemx} = \left(\frac{\mpr}{\me} \right)^{1/2}.
% \eqe
Thus, the ratio of the observed X-ray luminosities of the two populations is approximately given by
\eqb
 \frac{L_{\rm e}^{\rm obs}}{L_{\rm p}^{\rm obs}} &\approx& \frac{L_{\rm e}^{\rm syn}}{L_{\rm p}^{\rm syn}} \approx 
40 \ \frac{\leinj}{\lpinj}\frac{3-p}{2-p}\left(\frac{\gpmx}{10^6}\frac{R}{10^{15}} \frac{\ub}{10^2}\right)^{-1}.
\eqe
%\frac{L_{\rm e}^{\rm syn}}{L_{\rm p}^{\rm syn}} = 
% \frac{\leinj}{\lpinj}M^2\frac{F_{\rm p}}{F_{\rm e}}\frac{3-p}{3-s}\frac{\gemx^{3-s}}{\gpmx^{3-p}} = \\ \nonumber 
% & & \frac{\leinj}{\lpinj}M^2 G(p,s) \frac{\gemx}{\gpmx},
% \eqe
Assuming that $\frac{L_{\rm e}^{\rm obs}}{L_{\rm p}^{\rm obs}} = \chi \sim 0.1$,
the above equation gives a rough estimation for $\leinj$, which ensures that
the emission features of primary electrons in the X-rays will be `hidden' from the proton synchrotron component:
\eqb
\leinj \approx  (2.5\times10^{-5}) \frac{2-p}{3-p} \frac{\chi}{10^{-1}} \frac{\lpinj}{10^{-2}}
\frac{\gpmx}{10^6}\frac{R}{10^{15}} \frac{\ub}{10^2}.\eqe
For the exact values listed in Table~1 we find $\leinj \sim 4\times10^{-5} $.
\section[]{Simplified model for the system's response to variations of the source}
One of the factors that determines
the response of a particle/radiation system to the
variations of the source is the relation between 
the variability timescale of the source, which in our simulations
was taken equal to $1 \tcr$, and the minimum of the cooling and escape
timescales of particles.

Here we show through a smiplified model, that even in 
the non-physical case of sub-dynamical
cooling of particles,  i.e. $t_{\rm cool} << \tcr$, the photon 
distribution cannot track exactly the variations of the source.
For simplicity, we study the evolution of electrons and photons only. 
Let us assume that electrons, with number density $n$,
are being injected into a volume
with rate $Q$ and escape in a typical timescale $\teesc$, while
their characteristic cooling timescale is $\tecool$ ; energy losses
of electrons are treated as catastrophic. Photons, on the other hand,
are escaping within $\tcr$ and are being injected with a rate $\propto n/\tecool$.
Thus, the system is described by

\begin{figure}
\centering
\begin{tabular}{l}
\includegraphics[width=8cm, height=5cm]{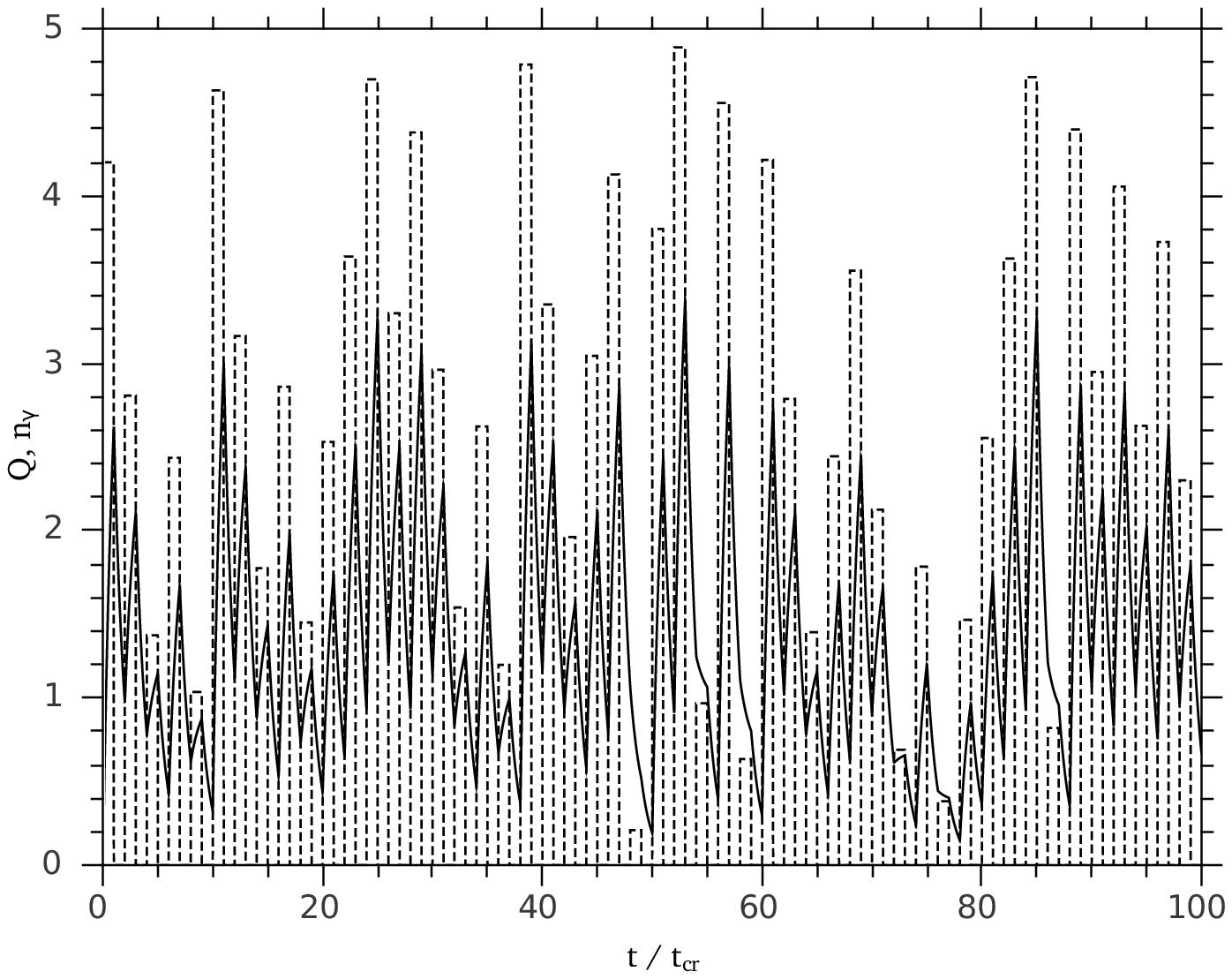} \\
\includegraphics[width=8cm, height=5cm]{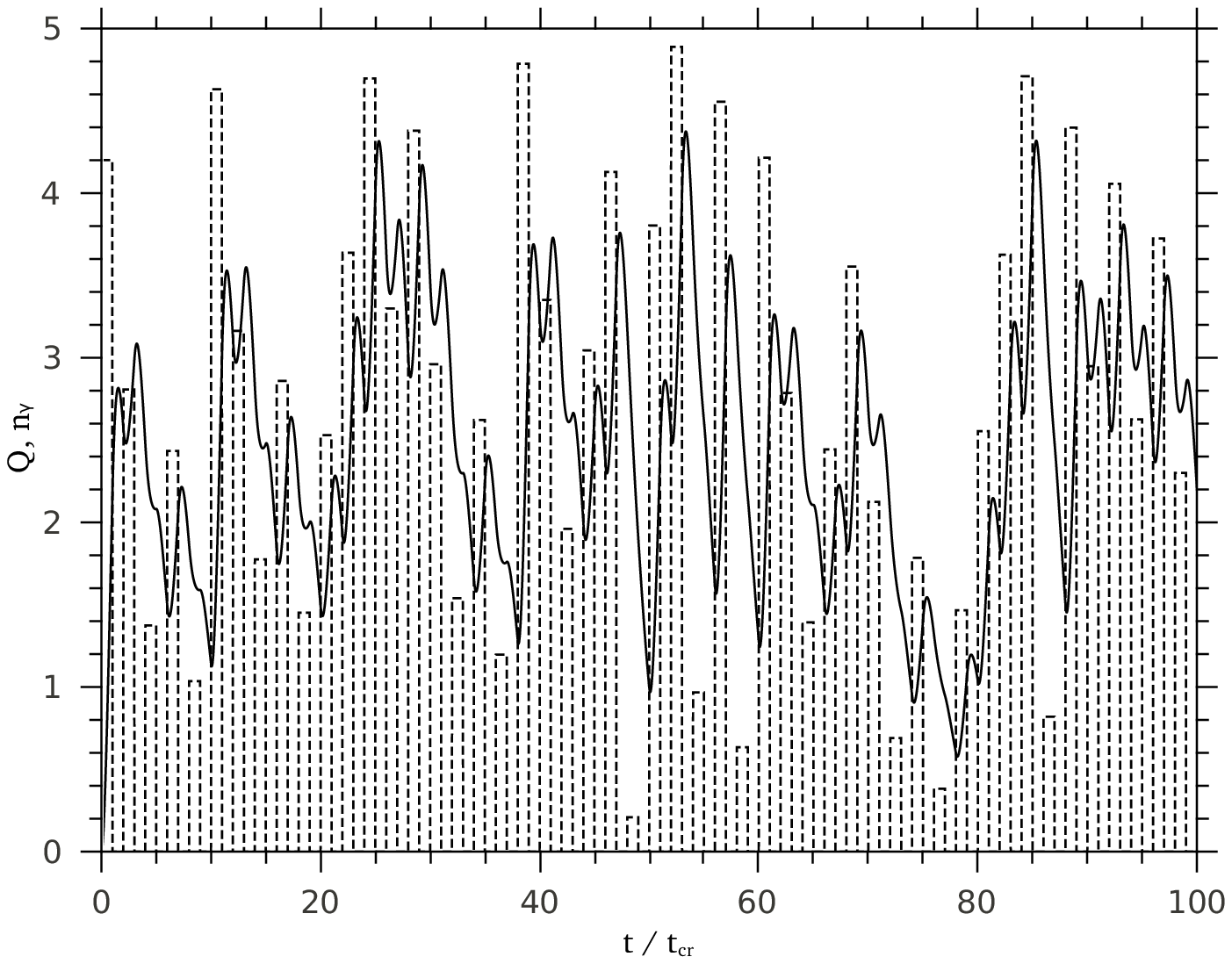} \\
\includegraphics[width=8cm, height=5cm]{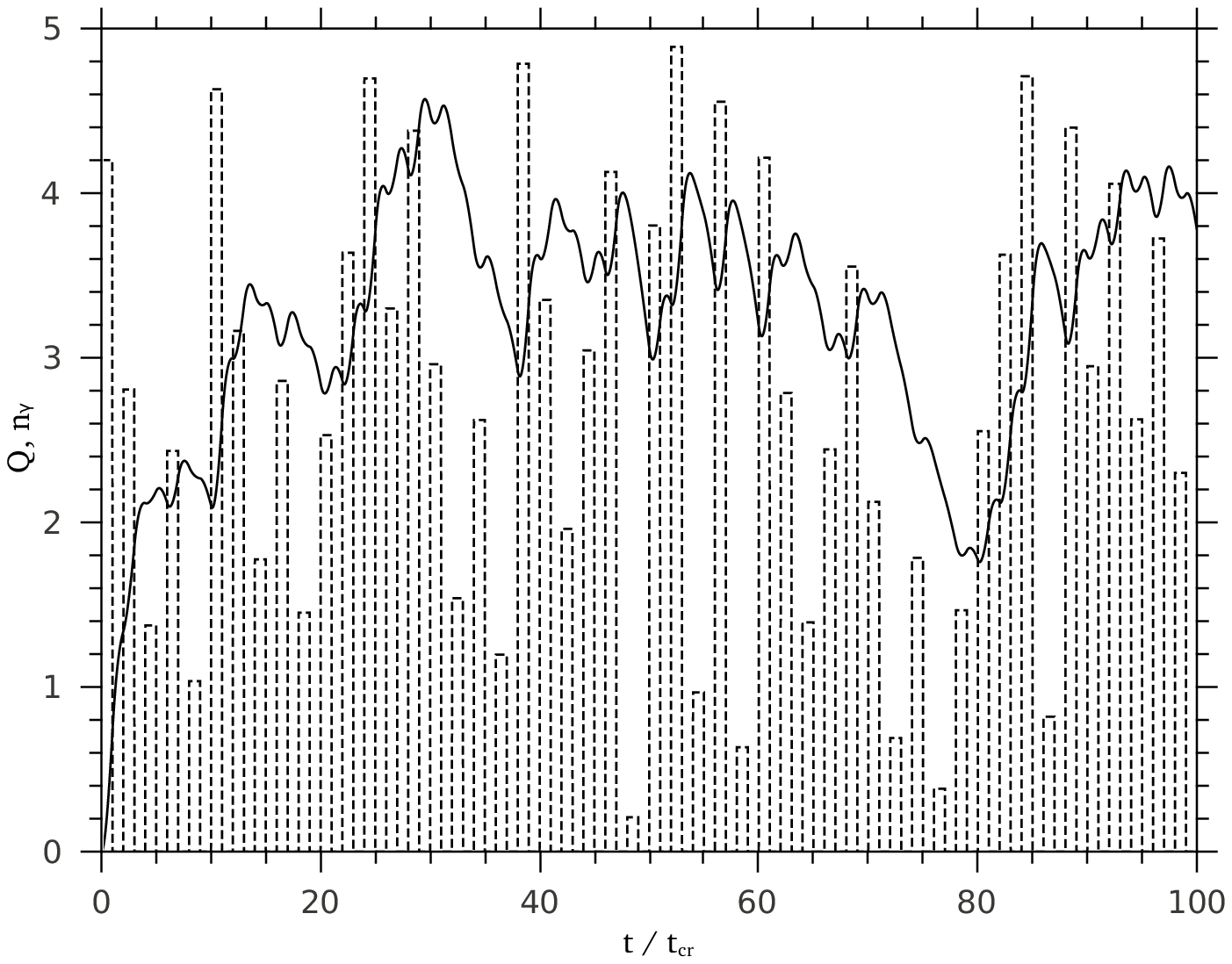}
\end{tabular}
\caption{Response of the photon distribution to the variations of electron injection for:
$\chi=100, \alpha=1$ (top panel), $\chi=0.1, \alpha=1$ (middle panel) and $\chi=0.1, \alpha=0.1$ 
(bottom panel). The injection function is depicted with dashed lines in all cases.}
\label{appendix-all}
 \end{figure}

\eqb
\frac{dn}{d\tau} + n(\alpha+\chi) & = & Q(\tau) \\
\frac{dn_{\gamma}}{d\tau} + n & = & J \chi n,
\eqe
where $\tau=t/\tcr$, $\alpha=\tcr/\teesc$, $\chi=\tcr/\tecool$ and $J$ is 
normalization constant. 
We model the electron injection function as a series of pulses with constant duration ($\delta t= 1 \tcr$)
and variable amplitude:
\eqb
Q(\tau) = \sum_{i=1,3,5}^{\rm N} Q_{\rm i-1} H(\tau-\tau_{\rm i-1})H(\tau_{\rm i}-\tau),
\eqe
where $H(x)$ is the step function and 
\eqb
Q_{\rm j}=Q_{\rm j-2}+r, \ Q_0=2 \ \textrm{and j}=2,4,6,...
\eqe 
and $r$ is a uniformly distributed random number in the range [-3,2]
while $\tau_{\rm i}=\tau_{\rm i-1}+\delta \tau$ and $\tau_0 =0$.
For initial conditions $n(0)=n_{\gamma}(0)=0$, the solution
for electrons and photons is given by:\\ \\
\textit{Electrons:}
\eqb
n=f_{\rm i-1}(\tau) \left\{ 
\begin{array} {ll}
 n(\tau_{\rm i-1})+\frac{Q_{i-1}}{\alpha+\chi}\frac{1-f_{\rm i-1}(\tau)}{f_{\rm i-1}(\tau)}, &  \textrm{i}=1,3,...\\
n(\tau_{\rm i-1}), &  \textrm{i}=2,4,...
\end{array} \right.
\eqe
where 
\eqb
f_{\rm i-1}(\tau)= e^{(\alpha+\chi)(\tau_{\rm i-1}-\tau)}
\eqe
and each branch is valid in the time interval $[\tau_{\rm i-1}, \tau_{\rm i}]$.\\ \\
\textit{Photons:} 
\eqb
n_{\gamma}& = & n_{\gamma}(\tau_{\rm i-1})E_{\rm i-1}(\tau) + J \chi \frac{Q_{\rm i-1}}{\alpha+\chi}\left(1-E_{\rm i-1}(\tau)\right) + \\ \nonumber
& &J \chi G_{\rm i-1}(\tau)\left(n(\tau_{\rm i-1})-\frac{Q_{\rm i-1}}{\alpha+\chi}\right) ,\ \textrm{i=1,3,5,...}
\eqe
where 
\eqb
G_{\rm i-1}(\tau) & \equiv &  \frac{\left(E_{\rm i-1}(\tau)\right)^{\alpha+\chi}-E_{\rm i-1}(\tau)}{1-\alpha-\chi} \quad \textrm{and}\\
E_{\rm i-1}(\tau) & \equiv & e^{\tau_{\rm i-1}-\tau}.
\eqe
The solution for $i=2,4,6$ is given below
\eqb
n_{\gamma}& =&  n_{\gamma}(\tau_{\rm i-1})E_{\rm i-1}(\tau) + J \chi 
\frac{n(\tau_{\rm i-1})}{1-\alpha-\chi}\cdot \\ \nonumber
& & \cdot\left[ \left(E_{\rm i-1}(\tau)\right)^{\alpha+\chi}-E_{\rm i-1}(\tau)\right].
\eqe
Figure \ref{appendix-all} shows the time-evolution of photons for three cases: 
(i) sub-dynamical cooling and fast particle escape with $\chi=100, \alpha=1$ (top panel), (ii)
slow cooling and fast escape with $\chi=0.1, \alpha=1$ (middle panel) and (iii) slow cooling and escape with $\chi=0.1, \alpha=0.1$ (bottom panel). 
The normalization constant in examples
(i)-(iii) was chosen to be $J=1, 20, 5$  respectively.
Note that even in the extreme case of ultra fast cooling (top panel) the photon lightcurve
does not have exactly the shape of the injection pulses. Thus, the power spectral density (PSD) of the
injection and photon time series will differ below some frequency $f_{\rm br}$. This effect will be even 
more prominent in examples (ii) and (iii), since the photon lightcurve becomes smoother and loses
the details in small timescales -- see middle and bottom panels in Fig.~\ref{appendix-all}.
In general, the temporal behaviour of the system including all physical processes, is
satisfactorily described by one
of the categories of the simplified model:
\begin{itemize}
 \item Fast cooling and escape, i.e. $t_{\rm cool} \lesssim t_{\rm esc} = \tcr$. The leptonic component of our models
belongs to this category, that corresponds to case (i) -- top panel in Fig.~\ref{appendix-all}. 
\item  Slow cooling and fast escape, i.e. $t_{\rm cool} >> t_{\rm esc}= \tcr$, which characterize
the proton distribution in all our three models. Only in the LHs model, protons at the high-energy tail
of the distribution have cooling timescales comparable to $\tcr$. 
 This category corresponds to case (ii) -- middle panel in 
 Fig.~\ref{appendix-all}. 
\item Slow cooling and escape, i.e. $t_{\rm cool} \approx t_{\rm esc} \gg \tcr$. 
For the fitting parameters adopted in the present
work, neither the proton nor 
the leptonic distibution falls into this category, which corresponds to case (iii) -- bottom panel in  
 Fig.~\ref{appendix-all}. 
\end{itemize}

\end{document}